\documentclass[conference]{IEEEtran}
\IEEEoverridecommandlockouts

\usepackage{mathtools}
\usepackage{amssymb}

\usepackage{xspace}
\usepackage[nolist]{acronym}
\usepackage[]{siunitx}
\usepackage{tabularx,colortbl}
\usepackage{booktabs}
\usepackage{multicol}
\usepackage{multirow}
\usepackage[nointegrals]{wasysym}
\usepackage{subcaption}
\usepackage{import}
\usepackage{enumitem}

\usepackage{placeins}
\usepackage{etoolbox}
\usepackage{nicefrac}

\usepackage{datatool, filecontents}
\DTLsetseparator{,}
\DTLloaddb{db-fm}{csv/results_fm.csv}
\DTLloaddb{db-doubledescription}{csv/results_dd.csv}
\DTLloaddb{db-backward}{csv/results_br.csv}
\DTLloaddb{db-adaptedIntegrationBounds}{csv/results_adapted_integration_bounds.csv}

\pretocmd{\section}{\FloatBarrier}{}{}

\usepackage{pgfplots}
\usetikzlibrary{arrows,arrows.meta,automata,backgrounds,positioning,intersections,shapes,decorations.pathreplacing,positioning, patterns, calc, shadows,shadows.blur,fit}

\pgfplotsset{compat=1.9}
\usepgfplotslibrary{fillbetween}

\usepackage[disable]{todonotes}
\usepackage{hyperref}

\newtheorem{runningexample}{Example}

\usepackage{pict2e,picture,graphicx}
\usepackage{wrapfig}
\usepackage{semantic}

\usepackage{caption}
\usepackage{url}
\usepackage{amsmath}
\usepackage{algorithm}
\usepackage{algpseudocode}

\usepackage{cite}
\usepackage{orcidlink}

\newenvironment{customalign}
 {$\!\aligned}
 {\endaligned$\\}
 
 \newtheorem{definition}{Definition}
 \newtheorem{lemma}{Lemma}

 \usepackage{dsfont}
 \usepackage{ifthen}
\begin{acronym}
\acro{ha}[HA]{hybrid automaton}
\acro{ta}[TA]{timed automaton}
\acro{swa}[SWA]{stopwatch automaton}
\acro{sa}[SA]{singular automaton}
\acro{ra}[RA]{rectangular automaton}
\acro{lha1}[LHA1]{linear hybrid automaton 1}
\acroplural{ha}[HA]{hybrid automata}
\acroplural{ta}[TA]{timed automata}
\acroplural{swa}[SWA]{stopwatch automata}
\acroplural{sa}[SA]{singular automata}
\acroplural{ra}[RA]{rectangular automata}
\acroplural{lha1}[LHA1]{linear hybrid automata 1}
\acro{rar}[RAR]{rectangular automata with random clocks}
\acro{sar}[SAR]{singular automata with random clocks}
\acro{sha}[SHA]{stochastic hybrid automaton}
\acroplural{sha}[SHA]{stochastic hybrid automata}
\acro{shs}[SHS]{stochastic hybrid system}
\acroplural{shs}[SHS]{stochastic hybrid systems}
\acro{hpng}[HPnG]{hybrid Petri net with general transitions}
\acroplural{hpng}[HPnGs]{hybrid Petri nets with general transitions}
\acro{mdp}[MDP]{Markov decision process}
\acroplural{mdp}[MDPs]{Markov decision processes}
\end{acronym}
\definecolor{niceblue}{RGB}{38 139 210}
\definecolor{niceyellow}{RGB}{181 137   0}

\definecolor{themecolor}{RGB}{0.15,0.38,0.61}
\definecolor{maincolor}{RGB}{0.15,0.38,0.61}

\definecolor{solBase01}{RGB}{88 110 117}
\definecolor{solBase1}{RGB}{147 161 161}
\definecolor{solMagenta}{RGB}{211  54 130}
\definecolor{color2}{named}{solMagenta}
\definecolor{solGreen}{RGB}{133 153   0}
\definecolor{color3}{named}{solGreen}
\definecolor{solYellow}{RGB}{181 137   0}
\definecolor{color4}{named}{solYellow}
\definecolor{dartmouthgreen}{rgb}{0.05, 0.5, 0.06}
\definecolor{color5}{named}{dartmouthgreen}
\definecolor{solBlue}{RGB}{38 139 210}
\definecolor{color6}{named}{solBlue}
\definecolor{davysgrey}{rgb}{0.33, 0.33, 0.33}
\definecolor{fancyblue}{rgb}{0.01, 0.28, 1.0}
\definecolor{amethyst}{rgb}{0.6, 0.4, 0.8}

\definecolor{boxcolor}{named}{color4}
\definecolor{refinementcolor}{named}{color2}

\definecolor{cinitloc}{named}{themecolor}
\definecolor{cpath1light}{named}{color6}
\definecolor{cpath1dark}{named}{fancyblue}
\definecolor{cpath2light}{named}{color3}
\definecolor{cpath2dark}{named}{color5}
\definecolor{crefseg}{named}{color2}
\definecolor{cinterseg}{named}{amethyst}
\definecolor{goaltrace1}{named}{cpath1dark}
\definecolor{goaltrace2}{named}{cpath2dark}
\newcommand{\realyst}{\textsc{RealySt}\xspace}
\newcommand{\hypro}{\textsc{HyPro}\xspace}

\newcommand{\prohver}{\textsc{ProHVer}\xspace}

\newcommand{\gsl}{\textsc{GSL}\xspace}

\newcommand\scalemath[2]{\scalebox{#1}{\mbox{\ensuremath{\displaystyle #2}}}}

\newcommand{\T}{\ensuremath{T}\xspace} %
\newcommand{\tmax}{\ensuremath{T_{\textsl{max}}}\xspace} %
\newcommand{\tint}{\ensuremath{t_{\textsl{int}}}\xspace} %
\newcommand{\estat}{\ensuremath{e_{\textsl{stat}}}\xspace}

\newcommand{\scheduler}{\ensuremath{\mathfrak{s}}\xspace}

\newcommand{\SchedulersProphetic}[1]{\ensuremath{\Schedulers}_{#1}\xspace} 
\newcommand{\Schedulers}{\ensuremath{\mathfrak{S}}\xspace}
\newcommand{\SchedulersProphMax}[1]{\ensuremath{\Schedulers_{#1}^{\textsl{max}}(\Goal, \tmax,\jumpmax)}\xspace}

\newcommand{\Distr}{\ensuremath{\mathit{Distr}}\xspace}

\newcommand{\sample}{\ensuremath{s}\xspace}

\newcommand{\States}[1]{\ensuremath{\mathcal{S}_{#1}}\xspace} %
\newcommand{\valRandom}{\ensuremath{\mu}\xspace}
\newcommand{\valSet}{\ensuremath{\mathcal{V}}\xspace} %
\newcommand{\valCRSet}{\ensuremath{\mathcal{P}}\xspace} %
\newcommand{\Goal}{\ensuremath{\smash{\mathcal{S}_\textsl{goal}}}\xspace}
\newcommand{\GoalE}{\ensuremath{\smash{\mathcal{S}_\textsl{goal}^{E}}}\xspace}
\newcommand{\GoalC}{\ensuremath{\smash{\mathcal{S}_\textsl{goal}^{C}}}\xspace}

\newcommand{\Goallocations}{\ensuremath{L^\textsl{goal}}\xspace}
\newcommand{\Goalvalset}{\ensuremath{\valSet^\textit{goal}}\xspace}

\newcommand{\Goalvali}{\ensuremath{\smash{\symb_{\mspace{1mu} i}^{\mspace{1mu} \textit{goal}}}}\xspace}
\newcommand{\Goalval}[1]{\ensuremath{\smash{\symb_{#1}^{\textit{goal}}}}\xspace}

\newcommand{\Goalreached}{\ensuremath{\smash{\mathcal{S}_\textsl{goal}^\textsl{reach}}}\xspace}

\newcommand{\symb}{\texttt{S}}
\newcommand{\symbType}[1]{\ensuremath{\symb_{#1}}\xspace} %
\newcommand{\repr}{\textit{repr}}

\newcommand{\trans}{f}

\newcommand{\comptime}{\ensuremath{t_\textsl{comp}}}

\newcommand{\scientific}[1]{\num[scientific-notation = true, exponent-product = \cdot]{#1}}
\newcommand{\rscientific}[1]{\num[scientific-notation = true, exponent-product = \cdot, round-mode = places, round-precision = 2]{#1}}
\newcommand{\samples}[1]{\scientific{#1}}
\newcommand{\errortable}[1]{\scriptsize\num[scientific-notation = true, exponent-product = \cdot, round-mode = places, round-precision = 3]{#1}}

\newcommand{\errortabletfont}[2]{\num[scientific-notation = true, exponent-product = \cdot, round-mode = places, round-precision = #1]{#2}}

\newcommand{\probab}[1]{\num[round-mode = places, round-precision = 6]{#1}}

\newcommand{\rintervals}[1]{\SI[round-mode = places, round-precision = 2]{#1}}
\newcommand{\rt}[1]{\SI[round-mode = places, round-precision = 2]{#1}{\hspace{-0.04cm}\second}}
\newcommand{\rtortimeout}[2]{\ifthenelse{\equal{#1}{#2}}{to}{\rt{#1}}}

\newcommand{\rtt}[2]{\SI[round-mode = places, round-precision = #1]{#2}{\second}}
\newcommand{\round}[1]{\SI[round-mode = places, round-precision = 2]{#1}{}}

\newcommand{\chull}[1]{\text{conv}(#1)}
\newcommand{\cone}[1]{\text{cone}(#1)}

\newcommand{\reachtree}{\ensuremath{\mathtt{R}}}

\newcommand{\Reals}{\ensuremath{\mathbb{R}}\xspace}

\newcommand{\Realsposzero}{\ensuremath{\Reals_{\geq0}}\xspace}

\newcommand{\Naturals}{\ensuremath{\mathbb{N}}\xspace}
\newcommand{\I}{\ensuremath{\mathbb{I}}\xspace}

\newcommand{\ARandom}{\ensuremath{\mathcal{R}}\xspace}
\newcommand{\RA}{\ensuremath{\ARandom}\xspace}
\newcommand{\RAC}{\ensuremath{\mathcal{C}}\xspace}

\newcommand{\RACU}{\ensuremath{\mathcal{C}_{u}}\xspace}

\newcommand{\tinyR}{{\scalemath{0.5}{R}}}
\newcommand{\tinyC}{{\scalemath{0.5}{C}}}
\newcommand{\tinyN}{{\scalemath{0.7}{N}}}
\newcommand{\tinyS}{{\scalemath{0.7}{S}}}

\newcommand{\Loc}{\ensuremath{\mathit{Loc}}\xspace} 
\newcommand{\VarCont}{\ensuremath{\mathit{Var}}\xspace}
\newcommand{\LabRandom}{\ensuremath{\mathit{Lab}}\xspace} 
\newcommand{\Edge}{\ensuremath{\mathit{Jump}\xspace}}

\newcommand{\Flow}{\ensuremath{\mathit{Flow}}\xspace}
\newcommand{\FlowCont}{\ensuremath{\Flow}\xspace}
\newcommand{\FlowRandom}[1]{\ensuremath{\Flow_{#1}}\xspace}
\newcommand{\Inv}{\ensuremath{\mathit{Inv}}\xspace}
\newcommand{\Init}{\ensuremath{\mathit{Init}}\xspace} 
\newcommand{\Event}{\ensuremath{\mathit{Event}}\xspace}

\newcommand{\valCont}{\ensuremath{\nu}\xspace} %

\newcommand{\CDF}{\ensuremath{\mathit{distr}}\xspace}
\newcommand{\CDFs}{\ensuremath{\mathbb{F}}\xspace}

\newcommand{\DFs}{\ensuremath{\mathbb{F}}\xspace}

\newcommand{\guard}{\mathit{guard}}
\newcommand{\reset}{\mathit{reset}}

\newcommand{\proj}[1]{\phantom{.}\!\!_{\vert #1}}

\newcommand{\clockr}{\ensuremath{r}\xspace}

\newcommand{\norc}{\ensuremath{\perp}\xspace}

\newcommand{\support}{\ensuremath{\mathit{supp}}\xspace}
\newcommand{\dimCont}{\ensuremath{{d_{\tinyC}}}\xspace}
\newcommand{\dimRandom}{\ensuremath{{d_{\tinyR}}}\xspace}
\newcommand{\dimRandomU}{\ensuremath{{d_{\tinyR}^{u}}}\xspace}
\newcommand{\dimCR}{\ensuremath{d}}
\newcommand{\identity}{\ensuremath{\mathit{id}}\xspace}
\newcommand{\jumpmax}{\ensuremath{\textsl{jmp}}\xspace} %
\newcommand{\prophecy}{\ensuremath{\kappa}\xspace} %
\newcommand{\Prophecies}[1]{\ensuremath{\mathcal{K}_{#1}}\xspace} %
\newcommand{\ftc}[1]{\ensuremath{\mathit{T^{+}_{#1}}}\xspace}
\newcommand{\depth}{\textit{depth}}
\newcommand{\source}{\textit{source}}
\newcommand{\target}{\textit{target}}
\newcommand{\fosr}[1]{\ensuremath{\mathit{D^{+}_{#1}}}\xspace}

\newcommand{\pathindices}{\ensuremath{N^{\textsl{goal}}}\xspace}

\newcommand{\Path}{\ensuremath{\pi}} %
\newcommand{\InitPaths}[1]{\ensuremath{\mathit{Runs}_{#1}^{\Init}}\xspace}
\newcommand{\last}{\textit{last}}
\newcommand{\hastate}{\ensuremath{\sigma}\xspace} %

\newcommand{\dur}{\mathit{dur}}
\newcommand{\Paths}[1]{\ensuremath{\mathit{Runs}_{#1}}\xspace}

\newcommand{\semanticsarrowrac}[1]{\ensuremath{\xrightarrow{#1}_{\scalemath{0.5}{\RAC}}}}
\newcommand{\semanticsarrowracu}[1]{\ensuremath{\xrightarrow{#1}_{\scalemath{0.5}{\RACU}}}}

\newcommand{\JumpChoices}{\ensuremath{\mathit{JumpChoices}}\xspace}
\newcommand{\TimeChoices}{\ensuremath{\mathit{TimeChoices}}\xspace}

\newcommand{\changed}[1]{{#1}}

\makeatletter
\DeclareRobustCommand{\Arrow}[1][]{%
\check@mathfonts
\if\relax\detokenize{#1}\relax
\settowidth{\dimen@}{$\m@th\rightarrow$}%
\else
\setlength{\dimen@}{#1}%
\fi
\sbox\z@{\usefont{U}{lasy}{m}{n}\symbol{41}}%
\begin{picture}(\dimen@,\ht\z@)
\roundcap
\put(\dimexpr\dimen@-.7\wd\z@,0){\usebox\z@}
\put(0,\fontdimen22\textfont2){\line(1,0){\dimen@}}
\end{picture}%
}
\makeatother

\makeatletter
\DeclareRobustCommand{\Barrow}[1][]{%
\check@mathfonts
\if\relax\detokenize{#1}\relax
\settowidth{\dimen@}{$\m@th\leftarrow$}%
\else
\setlength{\dimen@}{#1}%
\fi
\sbox\z@{\usefont{U}{lasy}{m}{n}\symbol{40}}%
\begin{picture}(\dimen@,\ht\z@)
\roundcap
\put(-1,0){\usebox\z@}%
\put(0,\fontdimen22\textfont2){\line(1,0){\dimen@}}%
\end{picture}%
}

\makeatother

\newcommand{\polyfwd}{\ensuremath{\mathcal{P}_{\!\!\scalemath{0.6}{\rightarrow}}}}
\newcommand{\polyfwdi}{\ensuremath{\mathcal{P}_{\!\!\scalemath{0.6}{\rightarrow}}^{i}}}
\newcommand{\polyfb}{\ensuremath{\mathcal{P}_{\!\!\scalemath{0.6}{\leftrightarrow}}}}

\newcommand{\hpoly}{H-polytope\xspace}

\newcommand{\hpolyhedra}{H-polyhedra\xspace}
\newcommand{\vpolyhedra}{V-polyhedra\xspace}

\tikzstyle{cone}=[draw = none, fill, fill opacity = 0.2]
\tikzstyle{border}=[very thick, line join = round]

\newcommand{\tikzbox}[1]{%
	\begin{tikzpicture}[scale=0.2, baseline=0mm,  thick]
		\draw[cone,#1] (0,0) rectangle (1.4,1.4);
		\draw[border, #1] (0,0) rectangle (1.4,1.4);
	\end{tikzpicture}%
}

\newcommand{\giga}{G}
\newcommand{\byte}{B}

\newcommand{\ebike}{e-bike\xspace}
\newcommand{\cebike}{EBIKE\xspace}
\newcommand{\car}{car\xspace}
\newcommand{\ccar}{CAR\xspace}

\newcommand{\DTLfetchsave}[5]{%
	\edtlgetrowforvalue{#1}{\dtlcolumnindex{#1}{#2}}{#3}%
	\dtlgetentryfromcurrentrow{#5}{\dtlcolumnindex{#1}{#4}}%
}
\input{tikzsettings}

\newif\ifarxiv\arxivtrue

\def\BibTeX{{\rm B\kern-.05em{\sc i\kern-.025em b}\kern-.08em
    T\kern-.1667em\lower.7ex\hbox{E}\kern-.125emX}}
\begin{document}
\begin{filecontents*}{csv/results_fm.csv}
table_index,probability,error_stat,error_infty,t_flowpipe,t_stochastic_processing,t_refinement_preprocessing,t_refinement,t_integration,runtime_0,runtime_1,runtime_2,runtime_21,runtime_22,runtime_23,runtime_24,runtime_25,runtime_26,runtime_27,runtime_28,max_constraint,filename
1,0.4636189079148219,0.0009682050487916415,0,0.164439125,0.028806401000000002,0.0789022,0.205778,0.0350228,0.514771,0.000178125,0.479533,0.164261,6.4501e-05,0.00058027,0.00013133,0.0789022,0.205778,0.0280303,0.0350228,56,../../scripts/results/realyst_fm_logging/car_0_ABC_rectangular_bR/car_0_ABC_rectangular_bR.log
2,0.6093029018877688,0.0008891698994034386,0,87.92909158600001,15.97635571,228.337,43.1242,25.9858,401.367,0.000391586,375.381,87.9287,0.00232717,0.00622544,0.0018031,228.337,43.1242,15.966,25.9858,724,../../scripts/results/realyst_fm_logging/car_1_ABC_rectangular_bR/car_1_ABC_rectangular_bR.log
3,0.3668866158861269,5.499435408969606e-05,0,0.159873007,0.044487739,0.0891607,0.316764,0.02243,0.633733,0.000210007,0.611054,0.159663,4.1824e-05,0.000232672,9.8643e-05,0.0891607,0.316764,0.0441146,0.02243,62,../../scripts/results/realyst_fm_logging/ebike_0_rectangular_bR/ebike_0_rectangular_bR.log
4,0.6869316365578081,0.000547088561059574,0,8.818257771999999,3.187484445,112.969,2879.91,16.1964,3021.09,0.000527772,3004.89,8.81773,0.000703945,0.00271589,0.00105461,112.969,2879.91,3.18301,16.1964,14457,../../scripts/results/realyst_fm_logging/ebike_1_rectangular_bR/ebike_1_rectangular_bR.log
5,0.4636189079148219,0.0009682050487916415,0,0.135993492,0.025332695,0.0555939,0.361945,0.0300075,0.609912,0.000144492,0.579732,0.135849,4.7063e-05,0.000380758,6.1874e-05,0.0555939,0.361945,0.024843,0.0300075,28,../../scripts/results/realyst_logging/car_0_ABC_rectangular_bR/car_0_ABC_rectangular_bR.log
6,0.6093029018877688,0.0008891698994034386,0,88.463189793,17.06017891,228.265,62.2625,26.9765,423.04,0.000389793,396.063,88.4628,0.00176963,0.0066576,0.00185168,228.265,62.2625,17.0499,26.9765,129,../../scripts/results/realyst_logging/car_1_ABC_rectangular_bR/car_1_ABC_rectangular_bR.log
7,0.3668866158861269,5.499435408969606e-05,0,0.153336906,0.043695624,0.0900258,0.58611,0.0185172,0.89293,0.000206906,0.874173,0.15313,4.1332e-05,0.00023893,0.000104162,0.0900258,0.58611,0.0433112,0.0185172,27,../../scripts/results/realyst_logging/ebike_0_rectangular_bR/ebike_0_rectangular_bR.log
8,0.6869316365578081,0.000547088561059574,0,8.617565551,3.264822456,108.594,20.1194,16.3874,156.987,0.000525551,140.599,8.61704,0.000442696,0.00253285,0.00104691,108.594,20.1194,3.2608,16.3874,130,../../scripts/results/realyst_logging/ebike_1_rectangular_bR/ebike_1_rectangular_bR.log
9,0.3114187220156135,0.0001659871244091629,0,0.146643978,0.027501984,0.0637114,0.123193,0.0190115,0.382046,0.000180978,0.362813,0.146463,6.4329e-05,0.000572828,0.000112227,0.0637114,0.123193,0.0267526,0.0190115,14,../../scripts/results/realyst_fm_logging/car_0_ABC_singular_bR/car_0_ABC_singular_bR.log
10,0.401543017557517,0.0001104801611287,0,37.615799026,10.92364145,96.2037,11.4063,8.90087,165.061,0.000399026,156.16,37.6154,0.00131954,0.00533813,0.00168378,96.2037,11.4063,10.9153,8.90087,32,../../scripts/results/realyst_fm_logging/car_1_ABC_singular_bR/car_1_ABC_singular_bR.log
11,0.1851606831447226,0.0003964709134959302,0,0.127941648,0.043853984,0.0470628,0.255338,0.0354953,0.510948,0.000222648,0.475186,0.127719,4.2273e-05,0.000247479,0.000119732,0.0470628,0.255338,0.0434445,0.0354953,16,../../scripts/results/realyst_fm_logging/ebike_0_singular_bR/ebike_0_singular_bR.log
12,0.3303389942663712,0.0002412775272057855,0,0.895485755,1.269689385,2.83553,3.67866,14.0568,22.7402,0.000512755,8.68279,0.894973,0.000611925,0.0033105,0.00120696,2.83553,3.67866,1.26456,14.0568,27,../../scripts/results/realyst_fm_logging/ebike_1_singular_bR/ebike_1_singular_bR.log
13,0.3114187220156135,0.0001659871244091629,0,0.134042538,0.021947444,0.0502661,0.265998,0.0151779,0.488453,0.000164538,0.473081,0.133878,4.4438e-05,0.000324669,6.0637e-05,0.0502661,0.265998,0.0215177,0.0151779,18,../../scripts/results/realyst_logging/car_0_ABC_singular_bR/car_0_ABC_singular_bR.log
14,0.401543017557517,0.0001104801611287,0,30.085434558,12.55495856,97.9661,41.8309,9.05623,191.504,0.000334558,182.447,30.0851,0.00132824,0.00540874,0.00172158,97.9661,41.8309,12.5465,9.05623,39,../../scripts/results/realyst_logging/car_1_ABC_singular_bR/car_1_ABC_singular_bR.log
15,0.1851606831447226,0.0003964709134959302,0,0.127020263,0.044165689,0.0480685,0.560626,0.0325504,0.813276,0.000293263,0.780396,0.126727,5.0424e-05,0.000314477,0.000164388,0.0480685,0.560626,0.0436364,0.0325504,17,../../scripts/results/realyst_logging/ebike_0_singular_bR/ebike_0_singular_bR.log
16,0.3303389942663712,0.0002412775272057855,0,0.897945297,1.368012376,2.77112,10.0301,14.1775,29.248,0.000548297,15.0698,0.897397,0.000557986,0.00300296,0.00112143,2.77112,10.0301,1.36333,14.1775,27,../../scripts/results/realyst_logging/ebike_1_singular_bR/ebike_1_singular_bR.log
\end{filecontents*}

\begin{filecontents*}{csv/results_dd.csv}
table_index,probability,error_stat,error_infty,t_flowpipe,t_stochastic_processing,t_refinement_preprocessing,t_refinement,t_integration,runtime_0,runtime_1,runtime_2,runtime_21,runtime_22,runtime_23,runtime_24,runtime_25,runtime_26,runtime_27,runtime_28,filename
1,0.4636189079148219,0.0009682050487916415,0,0.154354623,0.025236524,0.0670851,0.409904,0.0338303,0.692524,0.000180623,0.658462,0.154174,6.5082e-05,0.000350958,9.8984e-05,0.0670851,0.409904,0.0247215,0.0338303,../../scripts/results/realyst/car_0.log
2,0.6093029018877688,0.0008891698994034386,0,87.418888532,16.03192484,219.874,59.3124,25.1269,407.776,0.000388532,382.649,87.4185,0.00132487,0.00571762,0.00208235,219.874,59.3124,16.0228,25.1269,../../scripts/results/realyst/car_1.log
3,0.3668866158861269,5.499435408969606e-05,0,0.153289708,0.038445586,0.0793821,0.539187,0.0174041,0.82882,0.000224708,0.811136,0.153065,4.1973e-05,0.000257007,0.000100806,0.0793821,0.539187,0.0380458,0.0174041,../../scripts/results/realyst/ebike_0.log
4,0.6869316365578081,0.000547088561059574,0,8.267135847999999,2.9329923680000003,105.464,19.1312,15.8518,151.651,0.000505848,135.799,8.26663,0.000710248,0.00320873,0.00103339,105.464,19.1312,2.92804,15.8518,../../scripts/results/realyst/ebike_1.log
5,0.4636189079148219,0.0009682050487916415,0,0.177392346,0.031180178,0.0736983,0.420251,0.0322528,0.736896,0.000441346,0.70415,0.176951,6.7947e-05,0.000382095,0.000105336,0.0736983,0.420251,0.0306248,0.0322528,../../scripts/results/realyst_vtoh/car_0.log
6,0,0,0,0,0,0,0,0,36000,0,0,0,0,0,0,0,0,../../scripts/results/realyst_vtoh/car_1.log
7,0.3668866158861269,5.499435408969606e-05,0,0.251853681,0.052357884,0.129325,0.591378,0.0145716,1.04081,0.000555681,1.02563,0.251298,5.6369e-05,0.000361206,0.000152509,0.129325,0.591378,0.0517878,0.0145716,../../scripts/results/realyst_vtoh/ebike_0.log
8,0,0,0,0,0,0,0,0,36000,0,0,0,0,0,0,0,0,../../scripts/results/realyst_vtoh/ebike_1.log
9,0.4636189079148219,0.0009682050487916415,0,0.14270885700000002,0.015343619,0.0339118,0.396388,0.0330217,0.623255,0.000630857,0.589547,0.142078,6.2064e-05,0.000395389,0.000100366,0.0339118,0.396388,0.0147858,0.0330217,../../scripts/results/realyst_htov/car_0.log
10,0.6093029018877688,0.0008891698994034386,0,81.282388034,10.34117642,219.355,59.7992,22.9575,393.748,0.000388034,370.79,81.282,0.0013553,0.00550152,0.0020196,219.355,59.7992,10.3323,22.9575,../../scripts/results/realyst_htov/car_1.log
11,0.3668866158861269,5.499435408969606e-05,0,0.143935924,0.027841538,0.0693595,0.545828,0.017322,0.8057,0.000209924,0.788109,0.143726,4.1762e-05,0.000252376,9.77e-05,0.0693595,0.545828,0.0274497,0.017322,../../scripts/results/realyst_htov/ebike_0.log
12,0.6869316365578081,0.000547088561059574,0,8.274015614000001,1.680746386,103.79,19.1964,14.0179,146.964,0.000525614,132.946,8.27349,0.000706886,0.00441126,0.00104824,103.79,19.1964,1.67458,14.0179,../../scripts/results/realyst_htov/ebike_1.log
\end{filecontents*}

\begin{filecontents*}{csv/results_br.csv}
table_index,probability,error_stat,error_infty,t_flowpipe,t_stochastic_processing,t_refinement_preprocessing,t_refinement,t_integration,runtime_0,runtime_1,runtime_2,runtime_21,runtime_22,runtime_23,runtime_24,runtime_25,runtime_26,runtime_27,runtime_28,rectangular,filename
1,0.4636189079148219,0.0009682050487916415,0,0.144827633,0.0249594,0.0630796,0.40165,0.0335076,0.66945,0.000171633,0.635725,0.144656,5.759e-05,0.000315837,8.9373e-05,0.0630796,0.40165,0.0244966,0.0335076,1,../../scripts/results/realyst/car_0_ABC_rectangular_bR.log
2,0.6093029018877688,0.0008891698994034386,0,87.34890021700001,16.20658476,221.155,60.0215,25.5357,410.279,0.000400217,384.742,87.3485,0.00180497,0.00546813,0.00241166,221.155,60.0215,16.1969,25.5357,1,../../scripts/results/realyst/car_1_ABC_rectangular_bR.log
3,0.5603838433459739,2.04667728431503e-05,0,31663.300599004,5041.66051499,37395.0,95.313,1123.75,75319,0.000599004,74195.3,31663.3,0.0031102,0.00535145,0.00205334,37395,95.313,5041.65,1123.75,1,../../scripts/results/realyst/car_2_A_rectangular_bR.log
4,0.3668866158861269,5.499435408969606e-05,0,0.154463192,0.041878878999999994,0.0836606,0.564361,0.0179963,0.863654,0.000241192,0.845353,0.154222,4.4858e-05,0.000272434,0.000103987,0.0836606,0.564361,0.0414576,0.0179963,1,../../scripts/results/realyst/ebike_0_rectangular_bR.log
5,0.6869316365578081,0.000547088561059574,0,8.171500156,2.977905384,104.602,19.3547,16.1592,151.27,0.000510156,135.11,8.17099,0.000703884,0.00240606,0.00103544,104.602,19.3547,2.97376,16.1592,1,../../scripts/results/realyst/ebike_1_rectangular_bR.log
6,0.4636189079148219,0.0009682050487916415,0,0.149752047,0.01512731,0,0,0.0438641,0.2105,0.000183047,0.166396,0.149569,7.4384e-05,0.000374195,0.000102731,0,0,0.014576,0.0438641,1,../../scripts/results/realyst/car_0_ABC_rectangular_noBR.log
7,0.6093029018877688,0.0008891698994034386,0,87.222386323,6.4487966000000005,0,0,26.2763,119.956,0.000386323,93.6795,87.222,0.00159191,0.00532707,0.00203762,0,0,6.43984,26.2763,1,../../scripts/results/realyst/car_1_ABC_rectangular_noBR.log
8,0.5603838433459739,2.04667728431503e-05,0,33559.000615825,710.73363131,0,0,1112.95,35382.8,0.000615825,34269.8,33559,0.00384284,0.00569213,0.00209634,0,0,710.722,1112.95,1,../../scripts/results/realyst/car_2_A_rectangular_noBR.log
9,0.3668866158861269,5.499435408969606e-05,0,0.157703684,0.013640779,0,0,0.0211578,0.193767,0.000206684,0.172346,0.157497,6.0986e-05,0.000418641,0.000203552,0,0,0.0129576,0.0211578,1,../../scripts/results/realyst/ebike_0_rectangular_noBR.log
10,0.6869316365578081,0.000547088561059574,0,8.288510935000001,1.1182055160000002,0,0,16.0544,25.4644,0.000570935,9.40933,8.28794,0.000674436,0.00287974,0.00106134,0,0,1.11359,16.0544,1,../../scripts/results/realyst/ebike_1_rectangular_noBR.log
11,0.3114187220156135,0.0001659871244091629,0,0.135731612,0.024333689000000002,0.0545028,0.322657,0.0177902,0.556937,0.000165612,0.53893,0.135566,6.3815e-05,0.000316145,8.5329e-05,0.0545028,0.322657,0.0238684,0.0177902,0,../../scripts/results/realyst/car_0_ABC_singular_bR.log
12,0.401543017557517,0.0001104801611287,0,36.881171261,11.68950038,93.4976,39.5235,8.60392,190.207,0.000371261,181.603,36.8808,0.0012985,0.00464079,0.00166109,93.4976,39.5235,11.6819,8.60392,0,../../scripts/results/realyst/car_1_ABC_singular_bR.log
13,0.3430589779410709,0.0003328911778864901,0,418.644601638,3249.63878082,2166.77,79.3565,2333.75,8248.17,0.000601638,5914.42,418.644,0.00206351,0.00487635,0.00184096,2166.77,79.3565,3249.63,2333.75,0,../../scripts/results/realyst/car_2_A_singular_bR.log
14,0.1851606831447226,0.0003964709134959302,0,0.12482879300000001,0.03965728,0.0378536,0.489873,0.0308328,0.724114,0.000228793,0.692998,0.1246,4.1792e-05,0.00026516,0.000106128,0.0378536,0.489873,0.0392442,0.0308328,0,../../scripts/results/realyst/ebike_0_singular_bR.log
15,0.3303389942663712,0.0002412775272057855,0,0.851033223,1.198600112,2.62008,9.30336,14.0459,28.0216,0.000532223,13.975,0.850501,0.000556882,0.00277213,0.0011311,2.62008,9.30336,1.19414,14.0459,0,../../scripts/results/realyst/ebike_1_singular_bR.log
16,0.3114187220156135,0.0001659871244091629,0,0.135285759,0.009910722,0,0,0.0203756,0.16666,0.000172759,0.146059,0.135113,5.7239e-05,0.000306171,7.5992e-05,0,0,0.00947132,0.0203756,0,../../scripts/results/realyst/car_0_ABC_singular_noBR.log
17,0.401543017557517,0.0001104801611287,0,36.346445098,3.64452531,0,0,8.61963,48.6153,0.000645098,39.995,36.3458,0.00137096,0.00463784,0.00165651,0,0,3.63686,8.61963,0,../../scripts/results/realyst/car_1_ABC_singular_noBR.log
18,0.3430589779410709,0.0003328911778864901,0,417.721705406,250.23452453,0,0,2322.43,2990.4,0.000705406,667.971,417.721,0.00310987,0.00864271,0.00277195,0,0,250.22,2322.43,0,../../scripts/results/realyst/car_2_A_singular_noBR.log
19,0.1851606831447226,0.0003964709134959302,0,0.12772756,0.013070716999999999,0,0,0.0350153,0.176396,0.00023556,0.141089,0.127492,4.2613e-05,0.000285574,0.00010773,0,0,0.0126348,0.0350153,0,../../scripts/results/realyst/ebike_0_singular_noBR.log
20,0.3303389942663712,0.0002412775272057855,0,0.8650191939999999,0.587165437,0,0,13.9158,15.3704,0.000533194,1.45389,0.864486,0.000621607,0.00317752,0.00126331,0,0,0.582103,13.9158,0,../../scripts/results/realyst/ebike_1_singular_noBR.log
\end{filecontents*}

\begin{filecontents*}{csv/results_adapted_integration_bounds.csv}
table_index,probability,error_stat,error_infty,t_flowpipe,t_stochastic_processing,t_refinement_preprocessing,t_refinement,t_integration,runtime_0,runtime_1,runtime_2,runtime_21,runtime_22,runtime_23,runtime_24,runtime_25,runtime_26,runtime_27,runtime_28,bounding_box_lower_1,bounding_box_upper_1,bounding_box_lower_2,bounding_box_upper_2,bounding_box_lower_3,bounding_box_upper_3,bounding_box_lower_4,bounding_box_upper_4,bounding_box_lower_5,bounding_box_upper_5,filename
1,0.6069460216871232,0.0002286129443065118,0,84.9393717,14.5182933,216.664,55.9251,275.833,647.896,0.0005717,372.063,84.9388,0.0012184,0.0068881,0.0021868,216.664,55.9251,14.508,275.833,"0","16.6667","0","100","0","100","0","100","0","100",../../scripts/results/realyst/car_tint_100.log
2,0.6067748255891396,4.074020859424779e-05,0,92.7447599,13.9053313,214.364,54.908,133.53,509.473,0.0005599,375.942,92.7442,0.0017334,0.0069122,0.0021857,214.364,54.908,13.8945,133.53,"0","16.6667","0","26.8562","0","37.4299","0","75.4498","0","26.8562",../../scripts/results/realyst/car_tint_adapted.log
3,0.6093029018877688,0.0008891698994034386,0,92.73445530000001,14.873604400000001,214.321,54.3917,26.0664,402.409,0.0005553,376.342,92.7339,0.0012046,0.0067546,0.0023452,214.321,54.3917,14.8633,26.0664,"0","16.6667","0","100","0","100","0","100","0","100",../../scripts/results/realyst/car_tint_100_1e7.log
4,0.6075173161265921,0.000686090592558973,0,91.8079326,13.959828,212.956,55.0533,27.5403,401.336,0.0006326,373.795,91.8073,0.0011747,0.0064825,0.0021708,212.956,55.0533,13.95,27.5403,"0","16.6667","0","26.8562","0","37.4299","0","75.4498","0","26.8562",../../scripts/results/realyst/car_tint_adapted_1e7.log
\end{filecontents*}

\title{Scaling Up Reachability Analysis for Rectangular Automata with Random Clocks
\thanks{Supported by DFG project 471367371.}
}

\author{\IEEEauthorblockN{Jonas St\"{u}bbe}
\IEEEauthorblockA{\textit{University of M\"{u}nster} \\
Münster, Germany\\
jonas.stuebbe@uni-muenster.de \orcidlink{0000-0001-6586-2195}
}
\and
\IEEEauthorblockN{Anne Remke}
\IEEEauthorblockA{\textit{University of M\"{u}nster} \\
M\"{u}nster, Germany\\
anne.remke@uni-muenster.de \orcidlink{0000-0002-5912-4767}
}
\and
\IEEEauthorblockN{Erika \'{A}brah\'{a}m}
\IEEEauthorblockA{\textit{RWTH Aachen University} \\
Aachen, Germany\\
abraham@cs.rwth-aachen.de \orcidlink{0000-0002-5647-6134}
}
}

\maketitle

\begin{abstract}
This paper presents optimizations to improve the scalability of reachability analysis on a subclass of hybrid automata extended with stochasticity.
The optimizations target different components of the analysis, such as quantifier elimination for state set projection, and automated parameter selection during the numerical integration. Most importantly, whereas the original method combines forward and backward reachability, we show that the usage of backward reachability is optional for computing maximal reachability probabilities.
\end{abstract}


\section{Introduction}
Hybrid automata (HA) \cite{alur1995AlgorithmicAnalysisHybrid} are a popular formalism to model systems with mixed discrete-continuous behavior. This formalism has recently been extended with stochasticity in various forms~\cite{delicaris2023maximizing,Lygeros2010,AbateSHS2,bertrand2014StochasticTimedAutomata}; for an overview, we refer to~\cite{willemsen2023Qest}. This work considers systems whose continuous dynamics are intertwined with discrete events that occur after a (non-)deterministic or random amount of time.

Reachability analysis aims to compute the set of states that can be reached from predefined initial states of the system. This can be done with or without a predefined finite time and jump horizon, which then corresponds to bounded  or unbounded reachability analysis. Unbounded reachability is already undecidable for non-initialized singular automata, and hence also for non-initialized rectangular automata. However, time- and jump-bounded reachability is decidable for rectangular automata ~\cite{alur1995AlgorithmicAnalysisHybrid,frehse2005PHAVerAlgorithmicVerification}, which has been exploited in~\cite{delicaris2023maximizing} to compute maximum reachability probabilities under prophetic schedulers for rectangular automata with random clocks. 
The computation relies on polyhedra as  state-set representation and has been implemented within the tool \realyst~\cite{delicaris2024RealystToolPaper}, which relies on \hypro~\cite{schupp2017hypro} for flowpipe analysis and efficient operations on state sets. %

This paper improves the efficiency of time-bounded reachability analysis for rectangular automata with random clocks in several ways:

\begin{enumerate}[label=(\roman*)]
    \item Improving the efficiency of backward computation, required for computing schedulers that realize a maximum reachability probability (Sec. \ref{subsec:fm_qe}).
    \item Showing that forward analysis is sufficient for obtaining  maximum reachability probabilities (Sec. \ref{subsec:backwards_refinement}).
    \item Improving the numerical integration bounds in an automated way (Sec. \ref{subsec:dynamic_integration_bound}).
\end{enumerate}
Furthermore, we compare the efficiency of the approach from~\cite{delicaris2024RealystToolPaper} with the improvements proposed in this paper, using both a new and an existing case study. 

\paragraph{Related work.}

 CEGAR-style abstractions for probabilistic hybrid automata facilitate model checking by abstracting the system’s behavior through iterative refinements~\cite{franzle2011MeasurabilitySafetyVerification}. This enhances the applicability of timed automata by incorporating probabilistic elements and ensures that reachability remains decidable even when stochastic behavior is introduced by discrete probability distributions~\cite{sproston2000decidable,sproston2019Verification}. Other methods exclude nondeterminism ~\cite{bujorianu25,hu,bertrand2014StochasticTimedAutomata}  or apply discretization to random delays~\cite{zhang2012SafetyVerificationProbabilistic}. %
 Parametric reachability analysis offers a framework for exploring all reachable states by distinguishing between deterministic and stochastic events~\cite{gribaudo2016HybridPetriNets,ghasemieh2016survivability} for hybrid Petri nets with general transitions. %

Another approach is forward flowpipe construction, which incrementally computes the complete set of reachable states from the initial set, thereby producing the exact or approximate reachable state set depending on the  class of automata~\cite{frehse2005PHAVerAlgorithmicVerification}. %
The \hypro library \cite{schupp2017hypro} offers a suite of algorithms for flowpipe construction and state-set representations, utilizing structures such as polyhedra, alongside operations on them.
Specifically, Fourier-Motzkin quantifier elimination~\cite{schupp2019diss} is highly relevant for our approach but results in extensive computations.

\paragraph{Outline.}
Section \ref{sec:background} introduces the  foundations. Section \ref{sec:optimizations} presents  optimization strategies to improve the scalability of reachability analysis and evaluates their effectiveness on two case studies. Finally, Section \ref{sec:conclusion} concludes the paper. %

\section{Theoretical Background}\label{sec:background}
First, we introduce the symbolic representation of state sets in Section \ref{subsec:state-set representations}, which is used by \emph{rectangular automata with random clocks} (\emph{RAC}) \cite{delicaris2023maximizing} defined in Section \ref{subsec:model}. Next, we discuss how to resolve the nondeterminism inherent in RAC in Section \ref{subsec:max_sched}, and finally, we present their reachability analysis as proposed in \cite{delicaris2024Journal} in Section \ref{subsec:reachability_analysis}.
We present definitions from~\cite{delicaris2024Journal} syntacticaly adjusted for readability, while preserving the original meaning. The full definitions can be found in \ifarxiv{Appendix \ref{sec:appendix_definitions}}\else{\cite{stuebbe2025arxiv}}\fi.

\subsection{State-Set Representations and Conversion Methods}\label{subsec:state-set representations}
We recall some algebraic notions. A set $M \subseteq \Reals^{d}$ is convex iff $ \lambda x + (1-\lambda)y \in M $ for all $x,y \in M$ and $\lambda \in [0,1]\subseteq\Reals$. The \textit{convex hull of $M$} is the smallest convex set $\chull{M}\subseteq\Reals^d$ that contains $M$. The \textit{(convex) cone of $M$} is the smallest set $\cone{M}\subseteq\Reals^d$ that contains $M$ and for which $\alpha x+\beta y\in\cone{M}$ for all $x,y\in\cone{M}$ and $\alpha,\beta\in\Reals_{\geq 0}$. The \emph{Minkowski sum} of two sets $A$ and $B$ is $A\oplus B=\{a+b\,|\,a\in A\wedge b\in B\}$. Polyhedra can be represented by their defining halfspaces, or  as the convex hull of a finite set of points bloated by a cone.

\begin{definition}[H- and V-polyhedra \cite{ziegler2012lectures}]
A \emph{($d$-dimensional) H-polyhedron} is a pair $(A,b)$ with $A\in \Reals^{m\times d}$ and $b\in \Reals^m$ for some $m\in\Naturals$, which \emph{represents} the polyhedron $P(A, b) = \{ x\in \Reals^{d} \mid Ax \leq b \}$. A \emph{($d$-dimensional) V-polyhedron} is a pair  $(X,Y)\subseteq \Reals^{d}$, which \emph{represents} the polyhedron $P(X,Y) = \chull{X}\oplus\cone{Y}$.
\end{definition}

The two representations differ significantly in computational efficiency: operations that are easy for \hpolyhedra are often inefficient for \vpolyhedra, and vice versa. Thus, converting between both representations is crucial. \textit{Vertex enumeration} finds the V-polyhedron corresponding to an H-polyhedron, while the inverse problem is called \textit{facet enumeration}. Quickhull \cite{barber1996quickhull} addresses both problems, but has exponential  worst case complexity  in the dimension of the polyhedron with a base dependent on the number of vertices. Reachability analysis operates on bounded polyhedra, known as polytopes; therefore, we will refer to these representations as polytopes throughout the evaluation and can omit the empty cone $Y$ in the V-representation. During the analysis, \realyst leverages both H- and V-representations of polytopes, converting between them using Quickhull to perform operations efficiently.

\subsection{Model Definition}\label{subsec:model}
\emph{Rectangular conditions} refer to cross products of intervals from 
$\mathbb{I}=\{ [a, b], [a, \infty), (-\infty, b]\,|\,a, b \in \mathbb{Q} \}\cup\{(-\infty, \infty)\}$.
Given an ordered set of $d$ variables containing $x$ in the $i$th position, for any domain $D$ and $q = (q_1, \dots, q_d) \in D^d$, we use $q \proj{x}$ to refer to $q_i$. We denote componentwise comparison for $u,v\in D^d$, by $u\leq v$.

Rectangular automata (RA) are a special class of hybrid automata \cite{alur1995AlgorithmicAnalysisHybrid} (HA), whose behavior is specified by rectangular conditions. To aid later notation, the following definition requires, %
that the enabledness of discrete transitions (jumps) is exclusively defined by their guards. 

\begin{definition}[RA syntax]
    \label{def:RA}
    A \emph{rectangular automaton (RA)} is a tuple $\RA=(\Loc, \allowbreak\VarCont, \allowbreak\Inv,  \allowbreak\Init,\allowbreak\FlowCont,\Edge)$ with the following properties:
    
    \begin{itemize}
    \item  $\Loc$ is a finite nonempty set  of \emph{locations}.
    \item  $\VarCont$ is a finite ordered set  of $|\!\VarCont| = \dimCont$ \emph{(continuous) variables}. 
    \item $\Inv$, $\Init$ and $\FlowCont$ are functions, all of type $\Loc \rightarrow \I^{\dimCont}$, assigning an \emph{invariant}, \emph{initial states} resp. \emph{(flow) rates} to each location, satisfying (i) $\Init(\ell)\subseteq\Inv(\ell)$ for all $\ell\in\Loc$, (ii) $\Init(\ell)\not=\emptyset$ for some $\ell\in\Loc$, and (iii) $\FlowCont(\ell)\not=\emptyset$ for all $\ell\in\Loc$.
    \item  $\Edge \subseteq  \Loc \times \I^{\dimCont} \times (\I\cup \{\identity\})^{\dimCont} \times   \Loc$ is a finite set  of \emph{jumps} $e=(\ell,\allowbreak\guard,\allowbreak\reset,\allowbreak \ell')\in\Edge$ with \emph{source location} $\source(e)=\ell$, \emph{target location} $\target(e)=\ell'$, \emph{guard} $\guard(e)=\guard$, and \emph{reset} $\reset(e)=\reset$.
     We require for each jump $e=(\ell,\allowbreak\guard,\allowbreak\reset,\allowbreak \ell')\in\Edge$ that $\guard\subseteq\Inv(\ell)$ and for all $x\in\VarCont$,
     (i) if $\reset\proj{x} = \identity$ then $\Inv(\ell)\proj{x} \cap \guard\proj{x} \subseteq \Inv(\ell')\proj{x}$, and (ii) if $\reset\proj{x} \in \I$ then $\reset\proj{x} \subseteq \Inv(\ell')\proj{x}$.
\end{itemize}
For simplicity, in illustrations we sometimes omit components of invariants that are guaranteed by the semantics, i.e., those that always hold in the given location on all initial paths. $\RA$ is called \emph{nonblocking}, if for each $\ell\in\Loc$ and $\nu\in\Inv(\ell)$, either there is a jump $e\in\Edge$ with $\nu\in\guard(e)$, or there are $t\in\mathbb{R}_{>0}$ and $\textit{rate}\in \FlowCont(\ell)$ with $\nu+t\cdot \textit{rate}\in\Inv(\ell)$.
\end{definition}

\noindent Rectangular automata with random clocks (RAC) \cite{delicaris2024Journal} extend RA with stochasticity via random delays specified by (continuous) distributions.
For a function $\CDF: \mathbb{R}_{\geq 0} \to [0, 1] \subseteq \mathbb{R}$, we define its \emph{support} as $\support(\CDF) = \{v \in \mathbb{R}_{\geq 0} \mid \CDF(v) > 0\}$. 
We call $\CDF$ a \emph{(continuous) distribution} if it is absolutely continuous and satisfies $\int_{0}^{\infty} \CDF(v) \, dv = 1$. Let $\DFs$ denote the set of all distributions.

\begin{definition}[RAC syntax]
    \label{def:RAC}
    A \emph{rectangular automaton with random clocks (RAC)} is a tuple $\RAC = (\RA,\LabRandom,\allowbreak\Distr,\Event)$ 
    with the following properties:
    
    \begin{itemize}
    \item  $\RA=(\Loc, \allowbreak\VarCont, \allowbreak\Inv,  \allowbreak\Init,\allowbreak\FlowCont,\Edge)$ is a RA.

    \item $\LabRandom$ with $\LabRandom\cap\VarCont=\emptyset$ is a finite ordered set of $|\LabRandom| =\dimRandom$ \emph{random clocks}. 

    \item  $\Distr:\LabRandom\rightarrow\CDFs$ is a function mapping each random clock to a distribution.

     \item $\Event: \Edge \rightarrow (\LabRandom \ \cup \{\norc\})$ declares jumps $e\in\Edge$ as \emph{stochastic} if $\Event(e)\in\LabRandom$, and as \emph{nonstochastic} otherwise. For each stochastic jump $e\in \Edge$ we require $\guard(e)=\Reals^{\dimCont}$.

    \item The RA obtained from $\RA$ by removing all stochastic jumps from its edge set $\Edge$ is nonblocking.

\end{itemize}
We define $\FlowRandom{\RAC}:\LabRandom\rightarrow\{0,1\}^{\dimRandom}$, such that for each random clock $\clockr\in \LabRandom$ and each location $\ell \in \Loc$, $\FlowRandom{\RAC}(\ell)\proj{\clockr} = 1$ if and only if there is a stochastic jump $e \in \Edge$ with $\source(e)=\ell$ and $\Event(e)=\clockr$.
\end{definition}

\begin{figure*}[tb]
    \centering
\[
\begin{array}{c}
\\\\
\inference[\texttt{Rule}\textsubscript{$\mathit{Flow}$}]
{
  (\ell,\valCont, \valRandom, \sample) \in \States{\RAC} \quad
  t\in\Realsposzero \quad
  \textit{rate} \in  \FlowCont(\ell) \\
  \valCont'=\valCont+t\cdot \textit{rate} \in \Inv(\ell)\quad
  \valRandom' = \valRandom + t\cdot \FlowRandom{\RAC}(\ell) \leq \sample 
}{
    (\ell,\valCont, \valRandom, \sample) \semanticsarrowrac{t} (\ell,\valCont',\valRandom', \sample)
}
\\[0.75cm]
\inference[\texttt{Rule}\textsubscript{$\mathit{Jump}_{\tinyN}$}]
{
  (\ell,\valCont, \valRandom, \sample) \in \States{\RAC} \quad
  e=(\ell,\guard  , \reset , \ell') \in \Edge \quad
  \Event(e)=\norc \\
  \valCont \in \guard \quad
  \valCont'\in\reset(\valCont)
}{
    (\ell,\valCont, \valRandom, \sample) \semanticsarrowrac{e} (\ell',\valCont',\valRandom, \sample)
}
\\[0.75cm]
\inference[\texttt{Rule}\textsubscript{$\mathit{Jump}_{\tinyS}$}]
{
  (\ell,\valCont, \valRandom, \sample) \in \States{\RAC} \quad
  e=(\ell,\guard  , \reset , \ell') \in \Edge \quad
  \Event(e)=r \in \LabRandom \\
  \valCont'\in\reset(\valCont) \quad
  \valRandom\proj{\clockr} = \sample\proj{\clockr} \quad
  \valRandom'\proj{\clockr} = 0 \quad
  \sample'\proj{\clockr} \in \support(\Distr({\clockr})) \\
  \forall \clockr' \in \LabRandom \setminus \{\clockr\}. \
  \valRandom'\proj{\clockr'}=\valRandom\proj{\clockr'} \land \sample'\proj{\clockr'} = \sample\proj{\clockr'}
}{
    (\ell,\valCont, \valRandom, \sample) \semanticsarrowrac{e} (\ell',\valCont',\valRandom', \sample')
}
\end{array}
\]
    \caption{Operational semantics for RAC $\RAC= (\RA,\LabRandom,\allowbreak\Distr,\Event)$ with $\RA=(\Loc, \allowbreak\VarCont, \allowbreak\Inv,  \allowbreak\Init,\allowbreak\FlowCont,\Edge)$.
    }
    \label{fig:operationalsemanticsRA}
\end{figure*}

Random clocks  model stochastic durations. Each random clock $\clockr$ has an has an expiration time sampled from its distribution. Once the $\clockr$-labelled jump has been enabled for a total duration equal to this expiration time (possibly across multiple periods of enabledness), the jump is triggered.
Accordingly, a \emph{state} of $\RAC$ is a tuple $\hastate = (\ell, \valCont, \valRandom, \sample)$, which specifies the current location $\ell\in\Loc$, the \emph{valuation} $\valCont\in\Inv(\ell)$ of the continuous variables,  
the \emph{durations of enabledness} $\valRandom = (\valRandom_1, \dots, \valRandom_{\dimRandom})\in\Reals_{\geq 0}^{\dimRandom}$, and \emph{expiration times} $\sample = (\sample_1, \dots, \sample_{\dimRandom})\geq\valRandom$ with $\sample_i\in\support(\Distr(\clockr))$ for all $i$. Let $\States{\RAC}$ be the set of all states of $\RAC$.
When the stochastic component is not relevant, in the notation we sometimes omit the last two components.

The operational semantics for $\RAC$, as depicted in Figure \ref{fig:operationalsemanticsRA}, defines how its state can change by time evolution (rule \texttt{Rule}\textsubscript{$\mathit{Flow}$}), by taking a nonstochastic jump (rule \texttt{Rule}\textsubscript{$\mathit{Jump}_{\tinyN}$}), or by taking a stochastic jump (rule \texttt{Rule}\textsubscript{$\mathit{Jump}_{\tinyS}$}). Time evolution changes the valuation according to the flows, respecting the invariants. Nonstochastic jumps can be taken if their guards are enabled, and they change only the location and the valuation according to the jump's reset. When taking a jump $e\in\Edge$ with reset $\reset$ in valuation $\valCont$, the value of each variable $x\in\VarCont$ remains $\valCont(x)$ if $\reset\proj{x}=\identity$, and it is set to some value from $\reset\proj{x}\subseteq\I$ otherwise; we overload notation and write $\reset(\valCont)$ for the resulting valuation. Stochastic jumps with a label $\clockr$ can be taken if $\clockr$'s enabling duration reached its sampled value; besides resetting the continuous variables, a new value is sampled for $\clockr$ and its enabling duration is reset to $0$.

The applied reset for jumps and the applied rate for time steps are chosen nondeterministically, but can be uniquely determined from the states before and after the step.
If necessary, we annotate accordingly: $\hastate \semanticsarrowrac{e, \textit{res}} \hastate'$ and $\hastate \semanticsarrowrac{t, \textit{rate}} \hastate'$.

\begin{definition}[Run of $\RAC$~\cite{delicaris2024Journal}]
A (finite) \emph{run} of $\RAC$ is a sequence $\Path=(\hastate_0,a_1,\hastate_1,\allowbreak a_2,\ldots,\hastate_n)$ such that (i) $\hastate_i=(\ell_i,\valCont_i,\valRandom_i,\sample_i)\in\States{\RAC}$ for all $i=0,\ldots,n$, (ii) $\hastate_i\semanticsarrowrac{a_{i+1}}\hastate_{i+1}$ for all $i=0,\ldots,n-1$, (iii) if $a_i\in\Reals_{\geq 0}$ is a time step then $a_{i+1}\in\Edge$ is a jump for all $i=1,\ldots,n-1$, and (iv) if $a_n\in\Reals_{\geq 0}$ is a time step then at least one jump is enabled in $\hastate_n$.

We define $ | \Path | = n$ as the \emph{length} of $\Path$, $\dur(\Path)$ as the sum of all time step lengths in $\Path$ and $\last(\Path)=\hastate_n$. We call $\Path$ \emph{initial} if $\valCont_0\in\Init(\ell_0)$, $\sample_{0}\proj{\clockr} \in \support(\Distr({\clockr}))$ and  $\valRandom_{0}\proj{\clockr} = 0$ for all $\clockr \in \LabRandom$.
Let $\Paths{\RAC}$ ($\InitPaths{\RAC}$) be the set of all (initial) runs of $\RAC$.
\end{definition}

\subsection{Maximal Schedulers}\label{subsec:max_sched}
A RAC \(\RAC\) exhibits nondeterminism in three ways: (1) selecting an initial state from \(\mathit{InitialChoices}_{\RAC}\), (2) determining the duration and rate of a time step extending a run \(\Path\), as given by \(\TimeChoices_{\RAC}(\Path)\), and (3) choosing from a state \(\hastate\) a jump and a reset, as specified by \(\JumpChoices_{\RAC}(\hastate)\).

\noindent\begin{customalign}
\noindent\mathit{InitialChoices}_{\RAC}&=\{(\ell,\valCont) \in \Loc\times\Reals^{\dimCont} \mid \valCont\in\Init(\ell)\} \\
\mathit{TimeChoices}_{\RAC}(\Path)&=\{(t,\mathit{rate})\in \Realsposzero\times \Reals^\dimCont \\
&\,|\,\exists \hastate'\in\States{\RAC}.\ (\Path,(t,\textit{rate}), \hastate')\in\Paths{\RAC}\} \\
\mathit{JumpChoices}_{\RAC}(\Path) &=\{(e,\textit{res}){\in}\Edge\times(\Reals\cup\{id\})^{\dimCont} \\
&\,|\, \exists \hastate'\in\States{\RAC}.\ (\Path, (e,\textit{res}), \hastate') \in \Paths{\RAC}\}
\end{customalign}

We apply \emph{prophetic schedulers}~\cite{Pilch2021Optimizing} to resolve the nondeterminism present in $\RAC$. Prophetic schedulers have full information about the history and future expiration times of all random clocks, enabling them to resolve nondeterminism in a way that anticipates future events. 
This comprehensive knowledge is provided by \emph{prophecies} \cite{delicaris2024Journal}
$\prophecy : (\LabRandom\times\Naturals)\rightarrow\Reals_{\geq 0}$,
where $\prophecy(r, i)$ denotes the $i$th expiration time of the random variable $r \in \LabRandom$. Let $\Prophecies{\RAC}$ be the set of all prophecies for $\RAC$.

Prophetic schedulers are particularly suitable for worst-case analysis. By having complete foresight, they can systematically evaluate the potential outcomes and steer the system towards the most probable paths leading to a given state. This capability is crucial for safety-critical systems where understanding and mitigating worst-case scenarios is essential for ensuring reliability and safety.

\begin{definition}[Scheduler~\cite{delicaris2024Journal}]
 \label{def:Scheduler}
 \sloppy A \emph{(prophetic history-dependent) scheduler $\scheduler$ for $\RAC$} defines for each prophecy $\prophecy\in\Prophecies{\RAC}$ a function  $\allowbreak\scheduler_{\prophecy} 
 $%
 , such that for the empty path $\epsilon$, $\scheduler_{\prophecy}(\epsilon)\in\mathit{InitialChoices}_{\RAC}$ and $\scheduler_{\prophecy}(\Path) \in \JumpChoices_{\RAC}(\Path) \cup \TimeChoices_{\RAC}(\Path)$ for every initial run  $\Path$ of $\RAC$.
Let $\SchedulersProphetic{\RAC}$ be the set of all schedulers for $\RAC$.
 \end{definition}

For a set of locations $\Goallocations$ and a polyhedral set of valuations $\Goalvalset$, we define the \emph{goal states} as $\Goal= \{ ( \ell,\valCont,\valRandom,\sample )\in\States{\RAC} \mid \ell \in \Goallocations \wedge\valCont\in\Goalvalset\}$, unbounded in the stochastic dimensions. 

\begin{definition}[Run induced by scheduler $\scheduler_{\prophecy}$ for RAC $\RAC$~\cite{delicaris2024Journal}]
For a RAC $\RAC$, a scheduler $\scheduler$ for $\RAC$, a prophecy $\prophecy$, and a fixed $n\in\Naturals$, \emph{the run of length $n$ induced by $\scheduler_{\prophecy}$ in $\RAC$} is the unique initial run $\Path(\RAC,\scheduler_{\prophecy},n)=(\hastate_0,a_1,\ldots,a_n,\hastate_n)\in\InitPaths{\RAC}$ with $\hastate_0=\scheduler_{\prophecy}(\epsilon)$ and $a_{i+1}=\scheduler_{\prophecy}(\hastate_0,a_1,\ldots,a_i,\hastate_i)$ for $i=0,\ldots,n-1$. %

For $\tmax\in\Reals_{\geq 0}$, $\jumpmax\in\Naturals$ and $\Goal \subseteq \States{\RAC}$, we say that \emph{$\scheduler_{\prophecy}$ reaches $\Goal$ in $\RAC$ within bounds $(\tmax,\jumpmax)$} iff in the run $\Path(\RAC,\scheduler_{\prophecy},n)=\Path=(\hastate_0,a_1,\ldots,a_n,\hastate_n)\in\InitPaths{\RAC}$ of length $n$ induced by $\scheduler_{\prophecy}$ (i) the number jumps $|\{ i\in\{1,\dots,n\} \mid a_i \in \Edge \}|$ does not exceed the jump bound $\jumpmax$ and (ii) either $\hastate_n \in \Goal$ and $\dur(\Path)\leq \tmax$, or  $a_n=t$ is a time step and there exists $0\leq t'<t$ such that $(\hastate_0,a_1,\ldots,\hastate_{n-1},t',\hastate_n')\in\InitPaths{\RAC}$, $\hastate_n'\in\Goal$, and $\dur(\Path)-t+t'\leq\tmax$.

\end{definition}

We define the maximum reachability probability for prophetic schedulers in a RAC $\RAC$ following the approach outlined in~\cite{delicaris2024Journal}.

\begin{definition}[Prophetic maximum reachability probability~\cite{delicaris2024Journal}]
    \label{def:probability} %
Let $\RAC$ be a RAC with random clocks $\LabRandom$ and states $\States{\RAC}$. Assume $\Goal \subseteq \States{\RAC}$, $\tmax \in \Realsposzero$ and $\jumpmax\in\Naturals$. For each scheduler $\scheduler\in\SchedulersProphetic{\RAC}$, let $\Prophecies{}^\scheduler\subseteq\Reals_{\geq 0}^{\dimRandom \times \jumpmax}$ be the set of all $\jumpmax$-bounded prophecies $\prophecy$ for which $\scheduler_{\prophecy}$ reaches $\Goal$ in $\RAC$ within bounds $(\tmax,\jumpmax)$. Then, the \emph{prophetic maximum reachability probability to reach $\Goal$ in $\RAC$ within bounds $(\tmax,\jumpmax)$} is: 
\begin{equation*}\label{eq:scheduler}
p_{\RAC}^\textsl{max}(\Goal, \tmax,\jumpmax) 
= \max_{\scheduler \in \SchedulersProphetic{\RAC}} \Big( \int_{\Prophecies{}^\scheduler} G(\prophecy) \ d\prophecy \Big),
\end{equation*}
where  $G(\prophecy)= \prod_{\clockr\in\LabRandom} \prod_{i=0}^{\jumpmax} \Distr(\clockr)(\prophecy(\clockr,i))$. 
Let $\SchedulersProphMax{\RAC} \subseteq\SchedulersProphetic{\RAC}$ be the set of all schedulers that achieve this maximal probability.
\end{definition}

\subsection{Reachability Analysis}\label{subsec:reachability_analysis}

To analyse a RAC $\RAC$ with a jump bound $\jumpmax$, we transform it to the \emph{unrolled} RAC $\RACU$, in which no two stochastic jumps with the same label are taken. We do so by introducing $\jumpmax$ copies of each random label, unrolling the automaton to a loop-free tree up to the specified jump bound, and starting from the successor location of each stochastic $\clockr$-labelled jump we replace $\clockr$ by a fresh copy (see \cite{delicaris2024Journal}).
This approach ensures that each occurrence of a random event is uniquely represented by its own dimension in the state space. Thus, if the stochastic dimension of $\RAC$ is $\dimRandom$, then the stochastic dimension of $\RACU$ is $\dimRandomU = \dimRandom \times (\jumpmax + 1)$. Furthermore, we introduce a fresh clock $\T$ (with derivative $1$, never reset) to measure the global time, and to restrict the reachability computations to a fixed time bound $\tmax\in\Reals_{\geq 0}$.

The analysis starts with a forward reachability analysis phase for \RACU. In this phase, whenever a stochastic jump occurs, we omit the reset of its stopwatch $\clockr$. Instead, its current value is frozen to store it for later usage. This does not affect the future execution, since a new random clock is used to replace $\clockr$ in the remainder of the execution. %
\ifarxiv{Appendix \ref{sec:appendix}~}\else{\cite{stuebbe2025arxiv}~}\fi presents a semantics which never resets random clocks and is used in the following when we refer to the executions of $\RACU$.

The forward reachability analysis computes a reach tree, whose nodes store all states that are reachable in the unrolled RAC via initial paths within a bounded time duration and using a limited number of discrete jumps.
Starting from some initial state set, we iteratively compute time and jump successors, but at this phase we neglect the sample component of the states. 
Let $\dimCR=\dimCont+\dimRandomU$. For $(\valCont,\valRandom),(\valCont',\valRandom')\in\Reals^{\dimCR}$, let $(\ell,\valCont,\valRandom)\semanticsarrowracu{a}(\ell',\valCont',\valRandom')$ denote that $(\ell,\valCont,\valRandom,\sample)\semanticsarrowracu{a}(\ell',\valCont',\valRandom',\sample')$ for some $\sample,\sample'\in\Reals^{\dimRandomU}$.
Intuitively, in this phase we assume that all involved distributions have support on $\Reals_{\geq 0}$, and we consider their impact in the later phase of integration.

We use \emph{symbolic state sets} $\symb=(\ell,\valCRSet)\in \symbType{\RACU}=\Loc\times2^{\Reals^{\dimCR}}$ as a data structure to represent $\repr(\symb)=\{ (\ell, \valCont, \valRandom, \sample)\in\States{\RACU} \mid (\valCont,\valRandom) \in \valCRSet \}$. We define
the \emph{time successors of $\symb$} as $\ftc{}(\symb) = (\ell,\valCRSet')$ with 
$\valCRSet'=\{(\valCont',\valRandom')\in\Reals^{\dimCR}\,|\,\exists (\valCont,\valRandom)\in\valCRSet. \exists t\in\Reals_{\geq 0}.\ (\ell,\valCont,\valRandom)\semanticsarrowracu{t}(\ell,\valCont',\valRandom')\wedge \valCont'(\T)\leq \tmax\}$, and the
\emph{jump successors of $\symb$ for $e\in\Edge$} with $\source(e)=\ell$ as $\fosr{e}(\symb) = (\target(e),\valCRSet')$, where $\valCRSet'=\{(\valCont',\valRandom')\in\Reals^{\dimCR}\,|\,\exists (\valCont,\valRandom)\in\valCRSet.\ (\ell,\valCont,\valRandom)\semanticsarrowracu{e}(\target(e),\valCont',\valRandom')\}$.

\begin{definition}[Reach tree \cite{delicaris2024Journal}]
Let  $\RACU = (\RA,\LabRandom,\allowbreak\Distr,\Event)$ be an unrolled RAC
with $\RA = \allowbreak(\Loc, \allowbreak\VarCont, \allowbreak\Inv,  \allowbreak\Init,\allowbreak\FlowCont,\allowbreak\Edge)$ and $|\VarCont| = \dimCont$.
Assume a location $\ell$ of $\RACU$ with $\Init(\ell)\not=\emptyset$, a goal state set $\Goal$ of $\RACU$, a time bound $\tmax\in\Reals_{\geq 0}$, and a jump bound $\jumpmax\in\Naturals$.
A \emph{$(\tmax,\jumpmax)$-bounded reach tree for $\RACU$ from $\ell$ to $\Goal$} is an annotated tree $\reachtree = (N,E)$ with a nonempty finite set of \emph{nodes} $N\subseteq \Naturals\times(\Loc\times 2^{\Reals^{\dimCR}})$ and a set of \emph{edges} $ E = N \times  \Edge \times N$, such that:
\begin{itemize}

\item for each $n_1=(i_1,\symb_1)\in N$ and $n_2=(i_2,\symb_2)\in N\setminus\{n_1\}$ we have $i_1\not=i_2$, and if $(n_1,e,n_2)\in E$ for some $e\in\Edge$ then $i_1<i_2$;

\item there is a \emph{root} node $\textit{root}=(i_0,\ftc{}((\ell_0,\Init(\ell_0)))$ with $\textit{root}\not=n_2$ for all $(n_1,e,n_2)\in E$; 

\item for all $n\in N\setminus\{\textit{root}\}$  exactly one $(n_1,e,n_2)\in E$ exists with $n_2=n$; we define $\textit{parent}(n)=n_1$ and  $\textit{children}(n_1)=\{n_2\in N\,|\,\exists e\in\Edge.\ (n_1,e,n_2)\in E\}$;

\item using $\depth(\textit{root})=0$ and $\depth(n)=\depth(\textit{parent}(n))+1$ for all $n\in N\setminus\{\textit{root}\}$, we require for each $n_1=(i_1,\symb_1)\in N$, $\symb_1=(\ell_1,\valCRSet_1)$:
\begin{itemize}
\item if $\textit{depth}(n_1)=\jumpmax$ then $\textit{children}(n_1)=\emptyset$,
\item otherwise for each $e\in\Edge$
\begin{itemize}
\item if $\source(e)=\ell_1$ and $\ftc{}(\fosr{e}(\symb_1))=(\ell_2,\valCRSet_2)$ with $\valCRSet_2\not=\emptyset$, then there is a unique $(n_1,e,n_2)\in E$ with $n_2=(i_2,(\ell_2,\valCRSet_2))$, and for all $(n_1,e',n_2')\in E\setminus\{(n_1,e,n_2)\}$ it holds true that $e\not=e'$, and
\item otherwise $e\not=e'$ for all $(n_1,e',n_2)\in E$,
\end{itemize}
\end{itemize} 
\end{itemize} 
\end{definition}

\noindent
In a reach tree $\reachtree=(N,E)$, we define $\pathindices$ as the set of all indices $i\in\Naturals$ of nodes $(i,\symb_i) \in N$ with $\repr(\symb_i) \cap \Goal  \not = \emptyset$. Let $\Goalreached = \{ \symb_i \mid (i,\symb_i)\in N \wedge i \in \pathindices  \}$.

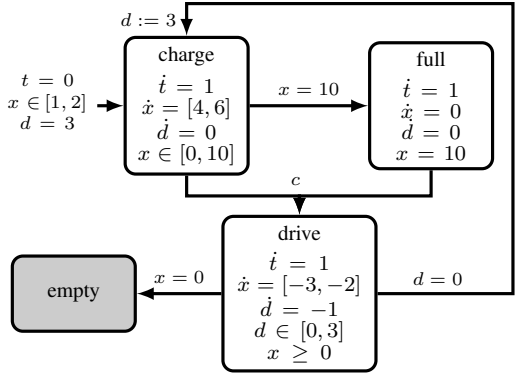
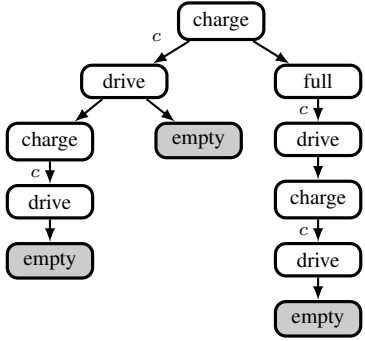
\begin{figure}[t]
\centering
\begin{subfigure}[b]{.99\linewidth}
\centering
					\begin{tikzpicture}[
			n/.style={draw, text width = 1.3cm, minimum height = 1.7cm, align = center, font= \footnotesize, rounded corners, very thick, execute at begin node=\setlength{\baselineskip}{8pt}%
			},
			nsmall/.style={draw, text width = 1.4cm, minimum height = 1cm, align = center, font= \footnotesize, rounded corners, very thick, execute at begin node=\setlength{\baselineskip}{8pt}%
			},
   nmediumsmall/.style={draw, text width = 1.8cm, minimum height = 1cm, align = center, font= \footnotesize, rounded corners, very thick, execute at begin node=\setlength{\baselineskip}{8pt}%
			},
			charging/.style={minimum height = 1.3cm
			},
			en/.style={draw=none, minimum height=0cm, font = \scriptsize, align = center},%
			c/.style={draw, fill, black, circle, inner sep=0, outer sep=0, minimum size=1mm},
			l/.style={anchor=west, inner sep=0, font=\footnotesize},
			]
			
			\node[nsmall,charging](chA) at (1.5,0) {
				$\text{charge}$	\\
				\vspace{0.125cm} 	
				$\dot{t}=1$\\
				$\dot{x}=[4,6]$\\
				$\dot{d}=0$\\
				$x\in[0,10]$
			};
			
			\node[nsmall,charging](noCh) at (4.75,0) {
				$\text{full}$	\\
				\vspace{0.125cm} 	
				$\dot{t}=1$\\
				$\dot{x}=0$\\
				$\dot{d}=0$\\
				$x=10$	};
			
			\node[nmediumsmall](dr) at ($(chA)-(-1.5,2.5)$) {
				$\text{drive}$\\
				\vspace{0.125cm} 	
				$\dot{t}=1$\\
				$\dot{x}=[-3,-2]$\\
				$\dot{d}=-1$\\
				$d\in[0,3]$\\
                $x \geq 0$
			};
			
			\node[nsmall, fill=maincolor!20!white](em) at  ($(chA)-(1.5,2.5)$)  {
				$\text{empty}$	
			};

			\node[en, text width=1.2cm,execute at begin node=\setlength{\baselineskip}{8pt}, anchor=north] (init) at (-0.35,0.54) {$t=0$\\$x\in${$\,[1,2]$}\\
				$d=3$ };
			\draw[-latex, very thick] ($(chA.west)-(0.35,0)$) to  ($(chA.west)+(0,0)$);

			\draw[-latex,  very thick ] (chA) to node[en, above] {$x=10$} (noCh);

\draw[-latex,  very thick] (chA.south) -- ($(chA)-(0,1.2)$) to node[en, xshift=0.7cm, yshift=0.2cm] {$c$} ($(chA)-(-1.5,1.2)$) -- (dr.north);
				\draw[-latex,  very thick] (noCh.south) -- ($(noCh)-(0,1.2)$) to ($(chA)-(-1.5,1.2)$) -- (dr.north);

			\draw[-latex,  very thick,] (dr) to node[en, above] {$x=0$} (em);
			
\draw[-latex,  very thick] ($(dr.east)$) node[en, xshift=0.8cm, yshift=0.2cm] {$d=0$} -- ($(0.25,0) + (noCh.east |- , |- dr.east)$) -- ($(0.25,0.5) + (noCh.east |- , |- noCh.north)$) -- ($(0.0,0.5) + (chA.north |- , |- noCh.north)$) to node[en, left] {$d:=3$} (chA);
			
\end{tikzpicture}%
            \caption{RAC $\RAC$ with $\LabRandom=\{c\}$.}\label{fig:runningex_automaton}
           
\end{subfigure}
\begin{subfigure}[b]{.99\linewidth}
\centering
\vspace*{0.5cm}
		  	\begin{tikzpicture}[
		scale=0.9, %
		node distance = 0.3cm and 0.5cm, %
		baseline,
		remember picture,
		n/.style={draw, text width = 0.9cm, text = black, %
			align = center, font= \footnotesize, rounded corners, very thick, execute at begin node=\setlength{\baselineskip}{7pt}%
		},
		en/.style={draw=none, minimum height=0cm, font = \scriptsize, align = center},%
		ref/.style={draw, circle, minimum width=2.0mm, inner sep=0, fill,refinementcolor},%
		c/.style={draw, fill, black, circle, inner sep=0, outer sep=0, minimum size=2mm},
		l/.style={anchor=west, inner sep=0, font=\footnotesize},
		]
		
		\node[n](chA0) at (0,0) {charge};
		
		\node[n,below left = 0.3cm and 0.7cm of chA0.south]  (dr0)  {drive};
		\node[n, below right = 0.3cm and 0.7cm of chA0.south] (noCh0) {full};
		
		\node[n, below left = 0.3cm and 0.4cm of dr0.south] (chA1) {charge};
		\node[n, below right = 0.3cm and 0.4cm of dr0.south, fill=maincolor!20!white] (emptyTrace0) {empty};
		\node[n, below = 0.3cm of noCh0.south] (dr0R) {drive};

		\node[n, below = of chA1.south] (dr1) {drive};

		\node[n, below = of dr0R.south] (chA1R) {charge};
		
		\node[n, below = of dr1.south, fill=maincolor!20!white] (emptyTrace1) {empty};

		\node[n, below = of chA1R.south] (dr1R) {drive};

		\node[n, below = of dr1R.south, fill=maincolor!20!white] (emptyTrace2) {empty};

		\draw[-latex, thick] (chA0) to node[en, left] {}  (noCh0);
		\draw[-latex, thick] (chA0) to node[en, left, yshift=0.2cm] {$c$}  (dr0);
		
		\draw[-latex, thick] (dr0) to node[en, left] {}  (chA1);
		\draw[-latex, thick] (dr0) to node[en, left] {}  (emptyTrace0);
	
		\draw[-latex, thick] (noCh0) to node[en, left] {$c$}  (dr0R);
		
		\draw[-latex, thick] (dr0R) to node[en, left] {}  (chA1R);	
	
		\draw[-latex, thick] (chA1) to node[en, left] {$c$}  (dr1);
	
		\draw[-latex, thick] (dr1) to node[en, left] {}  (emptyTrace1);
	
		\draw[-latex, thick] (chA1R) to node[en, left] {$c$}  (dr1R);
		
		\draw[-latex, thick] (dr1R) to node[en, left] {}  (emptyTrace2);
		
\end{tikzpicture}%
\caption{Reach tree structure.}\label{fig:runningex_reachtree}
\end{subfigure}
	\caption{RAC $\RAC$ for a simplified version of the CAR case study presented in~\cite{delicaris2024Journal,delicaris2023maximizing} and the structure of the corresponding reach tree $\reachtree$ for one cycle and goal state set \(\Goal = \{(\text{empty}, \valCont, \valRandom, \sample) \in \States{\RAC}\}\), omitting parts that do not reach the goal.}
	\label{fig:automaton_and_reachtree}
\end{figure}

\begin{runningexample}
Figure \ref{fig:runningex_automaton} illustrates  RAC $\RAC$ for a simplified version of the CAR case study presented in \cite{delicaris2024Journal,delicaris2023maximizing}, using a random clock \( c \) to model a battery charging process. Figure \ref{fig:runningex_reachtree} presents the structure of the corresponding reach tree $\reachtree$ for one cycle, \(\tmax = 100\), \(\jumpmax = 14\) and the goal state set \(\Goal = \{(\text{empty}, \valCont, \valRandom, \sample) \in \States{\RAC}\}\), indicating an empty battery. The analysis shows three distinct paths leading to the goal. Importantly, reaching the goal in the first cycle is impossible if the battery was fully charged. Let \( c_0 \) represent the valuation set of $c$, obtained from the flowpipe construction during the first cycle, and $c_1$ denote the corresponding valuation set in the second cycle.
The following valuation sets describe the three depicted paths, for details see \ifarxiv{Appendix \ref{sec:appendix}}\else{\cite{stuebbe2025arxiv}~}\fi:
\begin{enumerate}
   \item \( c_0 \in [\frac{2}{3}, 2.25], c_1 \in [0, 2.25] \) with the dependence \( c_0 + c_1 \leq 4.25 \).
\item \( c_0 \in [0, 2], c_1 \in [0, \infty] \).
\item \( c_0 \in [\frac{4}{3}, \frac{290}{3}], c_1 \in [0, 2] \) with the dependence \(c_0 + \frac{7}{3}c_1 \leq \frac{290}{3}\).
\end{enumerate}
\end{runningexample}
In previous work \cite{delicaris2024Journal}, the result of the forward analysis, as shown above, was complemented by a backward analysis. One of the contributions of this paper is to show that the backward analysis phase is optional for the computation of maximal reachability probabilities. However, if the schedulers, resolving the inherent nondeterminism are of interest, the backward analysis remains necessary. Due to space restrictions we omit a detailed description of the backward computation. Informally, the backward analysis intersects each state set $\symb_i\in\Goalreached$ in the reach tree with the goal states, and iteratively computes weakest preconditions up to the root, in order to identify those initial states from which the goal is reached along the respective reach tree branch. These weakest precondition computations use the methods of Gau\ss\ and Fourier-Motzkin (FM) to eliminate a variable $x$ from a set $C$ of linear real-arithmetic constraints, yielding a constraint set $C'$ that is equivalent to $\exists x.\ C$. If $x$ occurs in an equation $c\in C$, then the Gau\ss\ method transforms $c$ to the form $x=t$ for some linear term $t$, and substitutes $t$ for $x$ in all constraints in $C$ to achieve $C'$. Otherwise, if no equation contains $x$, then the FM method transforms each constraint that refers to $x$ to the form (L) $t\leq x$, $t<x$, or (U) $x\leq t$, $x<t$, depending on the sign of the coefficient of $x$. The L- and the U-constraints define lower respectively upper bounds on the value of $x$. Due to the density of the reals, there exists a value for $x$ satisfying $C$ iff none of the lower-upper-bound pairs $\ell\leq x$ and $x\leq u$ defines an empty interval $[\ell,u]$, or analogously $(\ell,u)$ if one of the bounds is strict. This requirement can be symbolically expressed as $\ell\leq u$ resp. $\ell<u$. The elimination of $x$ results in a new constraint set $C'$, containing all inequalities $\ell \le u$ (or $\ell < u$) derived from each lower–upper bound pair, along with the constraints in $C$ independent of $x$.

\section{Optimizations}\label{sec:optimizations}
We evaluate the introduced optimizations for unrolled RAC on a new case study, introduced in Section \ref{subsec:ebike}, 
and on the existing \ccar case study  which has previously been compared with the tool \prohver~\cite{HHHK13}  in~\cite{delicaris2024Journal,delicaris2023maximizing}.
Section \ref{subsec:fm_qe}  reduces constraint explosion during the time predecessor computations, necessary during backward computation. Section \ref{subsec:backwards_refinement}  shows that forward reachability analysis is sufficient to compute maximum reachability probabilities, if the actual schedulers are not of interest.
Section \ref{subsec:dynamic_integration_bound} identifies tighter bounds on the integration domain.

\subsection{Description of E-Bike case study}\label{subsec:ebike}

\def\ROUTEDISTANCE{20}\def\MAXCAPACITY{60}\def\SERVICETIME{15}
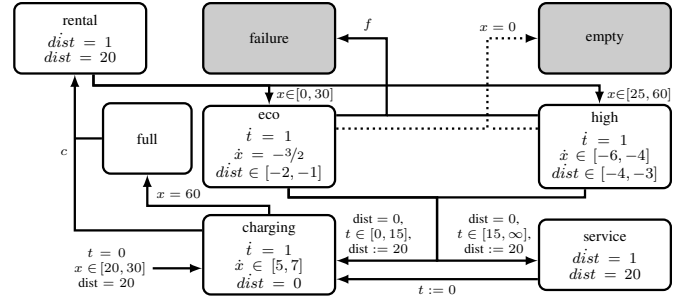
\begin{figure}[htb]
\def\singular{0}
\def\advancedFlowRepresentation{1}
\def\showInvariants{1}
   \scalebox{0.72}{%
   	\centering\def\advancedFlowRepresentation{0}
\def\showInvariants{0}

\begin{tikzpicture}[
	n/.style={draw, text width = 1.3cm, minimum height = 1.7cm, align = center, font= \footnotesize, rounded corners, very thick, execute at begin node=\setlength{\baselineskip}{8pt}%
	},
	nsmall/.style={draw, text width = 2.2cm, minimum height = 1cm, align = center, font= \footnotesize, rounded corners, very thick, execute at begin node=\setlength{\baselineskip}{8pt}%
	},
    ntiny/.style={draw, text width = 1.4cm, minimum height = 1cm, align = center, font= \footnotesize, rounded corners, very thick, execute at begin node=\setlength{\baselineskip}{8pt}%
	},
	charging/.style={minimum height = 1.3cm
	},
	en/.style={draw=none, minimum height=0cm, font = \scriptsize, align = center},%
	c/.style={draw, fill, black, circle, inner sep=0, outer sep=0, minimum size=1mm},
	l/.style={anchor=west, inner sep=0, font=\footnotesize},
 ]

	\node[nsmall,charging](eco) at (-2,0) {
		$\text{eco}$	\\
		\vspace{0.125cm} 	
		$\dot{t}=1$\\
		\if\singular1
		$\dot{x}=-1$\\
		$\dot{dist}=-1$\\
		\else
        \vspace{0.05cm}$\dot{x}=-\nicefrac{3}{2}$\\
		$\dot{dist}\in[-2,-1]$\\
		\fi
\if\advancedFlowRepresentation1
$\dot{f}=1$\\
$\dot{c}=0$\\
\else
\fi
	\if\showInvariants1
		$x\in[0,\MAXCAPACITY]$
		$dist\in[0,\ROUTEDISTANCE]$\\
		\else\fi
	};

	\node[nsmall,charging](high) at (4.2,0) {
	$\text{high}$	\\
	\vspace{0.125cm} 	
	$\dot{t}=1$\\
		\if\singular1
$\dot{x}=-5$\\
$\dot{dist}=-3$\\
\else
$\dot{x}\in[-6,-4]$\\
$\dot{dist}\in[-4,-3]$\\
\fi
\if\advancedFlowRepresentation1
$\dot{f}=1$\\
$\dot{c}=0$\\
\else
\fi
	\if\showInvariants1
	$x\in[0,\MAXCAPACITY]$
	$dist\in[0,\ROUTEDISTANCE]$\\
	\else\fi
};

	\node[nsmall,charging](service) at (4.2,-2) {
	$\text{service}$	\\
	\vspace{0.125cm} 	
	$\dot{dist}=1$\\
\if\advancedFlowRepresentation1
$\dot{t}=0$\\
$\dot{x}=0$\\
$\dot{f}=0$\\
$\dot{c}=0$\\
\else
\fi
	\if\showInvariants1
	$x\in[0,\MAXCAPACITY]$
	$dist=\ROUTEDISTANCE$\\
	\else
 	$dist=\ROUTEDISTANCE$\\
 \fi
};

	\node[nsmall,charging](charging) at (-2,-2) {
	$\text{charging}$	\\
	\vspace{0.125cm} 	
	$\dot{t}=1$\\
			\if\singular1
	$\dot{x}=5$\\
	\else
	$\dot{x}\in[5,7]$\\
	\fi
	$\dot{dist}=0$\\
\if\advancedFlowRepresentation1
$\dot{f}=1$\\
$\dot{c}=0$\\
\else
\fi
	\if\showInvariants1
$x\in[0,\MAXCAPACITY]$
$dist=\ROUTEDISTANCE$\\
\else\fi
};

	\node[nsmall,charging, fill=maincolor!20!white](empty) at (4.2,2) {
	$\text{empty}$	\\
	\vspace{0.125cm} 	
\if\advancedFlowRepresentation1
 $\dot{t}=0$\\
$\dot{x}=0$\\
$\dot{dist}=0$\\
$\dot{f}=0$\\
$\dot{c}=0$\\
\else
\fi
\if\showInvariants1
	$dist\in[0,\ROUTEDISTANCE]$
\else\fi
};
	
	\node[nsmall,charging, fill=maincolor!20!white](failure) at (-2,2) {
	$\text{failure}$	\\
	\vspace{0.125cm} 	
\if\advancedFlowRepresentation1
$\dot{t}=0$\\
$\dot{x}=0$\\
$\dot{dist}=0$\\
$\dot{f}=0$\\
$\dot{c}=0$\\
\else
\fi
	\if\showInvariants1
$x\in[0,\MAXCAPACITY]$
$dist\in[0,\ROUTEDISTANCE]$\\
\else\fi
};	

	\node[ntiny,charging](full) at (-4.25,0.15) {
	$\text{full}$	\\
	\vspace{0.125cm} 	
	\if\advancedFlowRepresentation1
 	$\dot{t}=0$\\
	$\dot{x}=0$\\
	$\dot{dist}=0$\\
	$\dot{f}=0$\\
	$\dot{c}=1$\\
	\else
	\fi
	\if\showInvariants1
$x\in[0,\MAXCAPACITY]$
$dist=\ROUTEDISTANCE$\\
	\else
\fi};

	\node[nsmall,charging](start) at ($(-2.25,0)+(failure.180)$) {
	$\text{rental}$	\\
	\vspace{0.125cm} 	
	$\dot{dist}=1$\\
\if\advancedFlowRepresentation1
$\dot{t}=0$\\
$\dot{x}=0$\\
$\dot{f}=0$\\
$\dot{c}=0$\\
\else
\fi
	\if\showInvariants1
$x\in[0,\MAXCAPACITY]$
$dist\in[0,\ROUTEDISTANCE]$\\
\else
$dist=\ROUTEDISTANCE$\\
\fi};

\draw[-latex,  very thick] (eco.295) -- ($(0,-0.15) + (eco.295 |- , |- high.south)$) -- ($(0,-0.15) + ( {$(eco)!0.5!(high)$} |- , |- high.south)$) -- ( {$(eco)!0.5!(high)$} |- 99 , 99 |- service.185) to node[en, above right=0.0cm and -0.7cm] {$\text{dist} = 0$,\\$t \in [\SERVICETIME, \infty]$,\\$\text{dist} := \ROUTEDISTANCE$} (service.185);

\draw[-latex,  very thick] (eco.295) -- ($(0,-0.15) + (eco.295 |- , |- high.south)$) -- ($(0,-0.15) + ( {$(eco)!0.5!(high)$} |- , |- high.south)$) -- ( {$(eco)!0.5!(high)$} |- 99 , 99 |- service.185) to node[en, above left=0.0cm  and -0.6cm] {$\text{dist} = 0$,\\$t \in [0, \SERVICETIME]$,\\$\text{dist} := \ROUTEDISTANCE$} (charging.355);

\draw[very thick] (high.245) -- ($(0,-0.15) + (high.245 |- , |- high.south)$) -- ($(0,-0.15) + ( {$(eco)!0.5!(high)$} |- , |- high.south)$);

\draw[-latex,  very thick] (service.200) to node[en, below=-0.025cm] {$t:=0$} ( charging.340 |- 99 , 99 |- service.200);
	
\draw[-latex,  very thick] (charging.north) -- ($(0,0.15) + (charging.north |- , |- charging.north)$) to node[en, above left=0cm and 0cm] {$x=\MAXCAPACITY$} ($(0,0.15) +( full.270 |- 99 , 99 |- charging.north)$) -- ( full.270 |- 99 , 99 |- full.270);

 \draw[-latex,  very thick] (charging.160) -- ($(-0.5,0) + (full.west |- , |- charging.160)$) to node[en, left] {$c$} ($(-0.5,0) + (full.west |- , |- start.250)$);

\draw[very thick] (full.west) -- ($(-0.5,0) + (full.west |- , |- full.west)$);

\draw[-latex,  very thick] (start.290) -- ($(0,-0.2) + (start.290)$) -- ($(0,-0.2) + (eco.north |- , |- start.290)$) to node[en, right] {$x{\in}[0,30]$} ($(eco.north) + (0,0)$);

\draw[-latex,  very thick] (start.290) -- ($(0,-0.2) + (start.290)$) -- ($(-0.1,-0.2) + (high.north |- , |- start.290)$) to node[en, right] {$x{\in}[25,60]$} ($(high.north) + (-0.1,0)$);

\draw[-latex,  very thick, dotted] (eco.15) -- ($(0,0) + ( {$(eco)!0.65!(high)$} |- , |- eco.15)$) -- ($(0,0) + ( {$(eco)!0.65!(high)$} |- , |- empty.180)$) to node[en, above left=0.0cm and -0.2cm] {$x=0$} (empty.180);
\draw[very thick, dotted] (high.165) -- ($(0,0) + ( {$(eco)!0.66!(high)$} |- , |- eco.15)$);

\draw[-latex,  very thick] (eco.25) -- ($(0,0) + ( {$(eco)!0.35!(high)$} |- , |- eco.25)$) -- ($(0,0) + ( {$(eco)!0.35!(high)$} |- , |- failure.0)$) to node[en, above right=0.0cm and -0.1cm] {$f$} (failure.0);

\draw[very thick] (high.155) -- ($(0,0) + ( {$(eco)!0.35!(high)$} |- , |- eco.25)$);
	
\node[en, text width=1.2cm,execute at begin node=\setlength{\baselineskip}{8pt}, anchor=north] (init) at (-5,-1.75) {$t=0$\\
	$x\in${$\,[20, 30]$}\\
	$\text{dist}=\ROUTEDISTANCE$
    };
\draw[-latex,  very thick] ($(0.1,-0.25) + ( full.south |- 99 , 99 |- charging.180)$) to node[en, above] {} ($(0,-0.25) + (charging.180)$);
	
\end{tikzpicture}
   }
	\caption{\cebike model with cycles modeled as RAC.}
	\label{fig:ebike_casestudy}
\end{figure}

We model an \ebike rental service and focus on the state-of-charge of their batteries,  modeled by variable $x$. A simplified RAC for this case study is shown in Figure \ref{fig:ebike_casestudy}; the full RAC is given in \ifarxiv{Appendix \ref{sec:appendix_casestudy}}\else{\cite{stuebbe2025arxiv}~}\fi. The \cebike model features several aspects, not present in the \ccar model, as follows: (1) multiple goal locations for reachability computations, (2) resets of continuous variables, (3) multiple rectangular flows per location. Especially (2) and (3) lead to specific scalability challenges, which do not occur in the \car model.

An \ebike operates either in mode \textit{eco} or in \textit{high}, with power consumption \(\dot{x}=-\frac{3}{2}\)  and  $\dot{x}\in [-6,-4]$, respectively. %
The rate at which the remaining distance \(\textit{dist}\) decreases is chosen  from a different interval in every location. The distance per rental is set to $20$  for simplicity,  which puts the focus of the model on battery consumption and charging rates. 
Charging times between rentals are determined by a random clock \(c\) following a folded normal distribution \(F_Y(\mu=6, \sigma=3)\). The charging rate \(\dot{x}\) is selected  from an interval. Battery failures during driving are modeled by random clock \(f\) with an exponential distribution \(\text{Exp}(\lambda=0.025)\), where longer driving times without service increase failure probability.

Locations in the automaton model different phases of a rental cycle: the location \textit{charging} models the beginning of a rental; \textit{eco} and \textit{high} are operating modes during cycling. When leaving \textit{rental}, the operating mode is selected based on the state-of-charge of the battery  \(x\).  If the battery is depleted, the automaton moves to \textit{empty}.
One rental cycle ends when \textit{dist} reaches $0$ and locations \textit{eco} or \textit{high} are left for locations   \textit{charging} or \textit{service}. 
Once in  location \textit{charging}, this process is either completed and location \textit{full} is entered, or the \ebike is rented with a state-of-charge that depends on the previous rental and the time spent in location \textit{charging}. Entering location \textit{rental} resets the random clock $c$. Location \textit{service} is entered if the time since the last service exceeds  threshold $t=15$, which resets the random clock $f$ modeling a failure of the battery. Note that this behavior is not directly visible in Figure \ref{fig:ebike_casestudy}, as the figure depicts the original, non-unrolled automaton. In the corresponding unrolled RAC, however, resetting the clocks $f$ and $c$ corresponds to introducing new independent random variables $f_i$ and $c_i$, respectively.
The maximum probability of battery depletion or failure is computed in the following. This corresponds to a worst case analysis of the the probability that a \ebike ride is unsuccessfull. 

\subsection{Parameter settings} The presented results are generated using the tool \realyst \footnote{See \url{https://go.uni-muenster.de/realyst} for tool and model files.}. 
We additionally adapt both case studies to singular variants by fixing the rectangular intervals to their mid-points; for the CAR case study, we consider variants with charging types A and ABC.

For the \cebike case study, we  define the goal set as  \(\GoalE = \{(\ell, \valCont, \valRandom, \sample) \in\States{\RAC} \mid \ell \in \{\text{empty}, \text{failure}\}\} \), covering all states with an empty battery and all states where the \ebike failed. 
We limit the evaluation of the \cebike case study by restricting the number of rental cycles (\(\texttt{\#}\text{cycles} \in \{0,1\}\), where $0$ corresponds to the first rental cycle).
The time and jump bounds are chosen as \(\tmax^{E} = (\texttt{\#}\text{cycles} + 1) \cdot 32\) and \(\jumpmax^{E} = (\texttt{\#}\text{cycles} + 1) \cdot 5\).

For the \ccar case study, we use the charging type configurations A and ABC, along with the same parameters as in \cite{delicaris2024Journal}, which we recall below for completeness.  The goal set is defined as  \(\GoalC = \{(\text{empty}, \valCont, \valRandom, \sample) \in\States{\RAC}\}\), indicating an empty battery. The time and jump bounds are defined as \(\tmax^{C} = (\texttt{\#}\text{cycles} + 1) \cdot 10\) and \(\jumpmax^{C} = (\texttt{\#}\text{cycles} + 1) \cdot 7\).
 
If not indicated otherwise, numerical integration uses sample sizes: \samples{1e5}, \samples{1e7}, \samples{1e9} for \ccar with 0, 1, and 2 cycles, respectively, and \samples{1e5}, \samples{1e7} for \cebike with 0 and 1 cycles, respectively. All experiments were conducted on a machine equipped with a 12th Gen Intel\smash{\textsuperscript{\textregistered}} Core\textsuperscript{\texttrademark} i7-1255U processor running at 10 \(\times\) \SI{1.70}{\giga\hertz} with \SI{64}{\giga\byte} of RAM.

\subsection{Fourier-Motzkin Quantifier Elimination}\label{subsec:fm_qe}
The methods explained in Section \ref{subsec:reachability_analysis} are implemented in the tool \realyst~\cite{delicaris2024RealystToolPaper}. They apply backward refinement and iteratively eliminate variables with \emph{Fourier-Motzkin},
combining lower and upper bounds and  generating  \( \mathcal{O}(n^2) \) constraints from \( n \) input constraints in the worst case.

\subsubsection{Checking for redundant constraints}
\begin{table}[t]
\centering

	\centering
	\scriptsize
    \captionof{table}{Computation time of \realyst and max. number of constraints N for FM($^{+}$) during backward refinement.}
	\label{tab:both_fm_full}
	\newcolumntype{Y}{>{\centering\arraybackslash}X}
	\renewcommand{\arraystretch}{1}
	\begin{tabularx}{\linewidth}{p{0.25cm}p{0.55cm}p{0.55cm}YYYY}
		\toprule
        \multicolumn{3}{c}{} & \multicolumn{2}{c}{\ccar \texttt{ABC}
        }&\multicolumn{2}{c}{\cebike}\\
        \cmidrule(lr){4-5}\cmidrule(lr){6-7}
		\multicolumn{3}{c}{cycles} & $0$  & $1$  & $0$  & $1$ \\
		\midrule
		\multirow{4}{*}{\rotatebox{90}{rectangular\phantom{xx}}}
		& \multirow{2}{*}{FM} & N
        &\DTLfetch{db-fm}{table_index}{1}{max_constraint}
        &\DTLfetch{db-fm}{table_index}{2}{max_constraint}
        &\DTLfetch{db-fm}{table_index}{3}{max_constraint}
        &\DTLfetch{db-fm}{table_index}{4}{max_constraint}\\
        \cmidrule(lr){3-7}
        & & {$\comptime$} 
        &\DTLfetchsave{db-fm}{table_index}{1}{runtime_0}{\runtimeTmp}\rt{\runtimeTmp}
        &\DTLfetchsave{db-fm}{table_index}{2}{runtime_0}{\runtimeTmp}\rt{\runtimeTmp}
        &\DTLfetchsave{db-fm}{table_index}{3}{runtime_0}{\runtimeTmp}\rt{\runtimeTmp}
        &\DTLfetchsave{db-fm}{table_index}{4}{runtime_0}{\runtimeTmp}\rt{\runtimeTmp}\\
        \cmidrule(lr){2-7}
        & \multirow{2}{*}{FM$^{+}$} & N 
        &\DTLfetch{db-fm}{table_index}{5}{max_constraint}
        &\DTLfetch{db-fm}{table_index}{6}{max_constraint}
        &\DTLfetch{db-fm}{table_index}{7}{max_constraint}
        &\DTLfetch{db-fm}{table_index}{8}{max_constraint}\\
         \cmidrule(lr){3-7}
        & & {$\comptime$}
        &\DTLfetchsave{db-fm}{table_index}{5}{runtime_0}{\runtimeTmp}\rt{\runtimeTmp}
        &\DTLfetchsave{db-fm}{table_index}{6}{runtime_0}{\runtimeTmp}\rt{\runtimeTmp}
        &\DTLfetchsave{db-fm}{table_index}{7}{runtime_0}{\runtimeTmp}\rt{\runtimeTmp}
        &\DTLfetchsave{db-fm}{table_index}{8}{runtime_0}{\runtimeTmp}\rt{\runtimeTmp}\\
        \midrule
        \multirow{4}{*}{\rotatebox{90}{singular \phantom{xx}}}
        & \multirow{2}{*}{FM} & N
        &\DTLfetch{db-fm}{table_index}{9}{max_constraint}
        &\DTLfetch{db-fm}{table_index}{10}{max_constraint}
        &\DTLfetch{db-fm}{table_index}{11}{max_constraint}
        &\DTLfetch{db-fm}{table_index}{12}{max_constraint}\\
        \cmidrule(lr){3-7}
        & & {$\comptime$}
        &\DTLfetchsave{db-fm}{table_index}{9}{runtime_0}{\runtimeTmp}\rt{\runtimeTmp}
        &\DTLfetchsave{db-fm}{table_index}{10}{runtime_0}{\runtimeTmp}\rt{\runtimeTmp}
        &\DTLfetchsave{db-fm}{table_index}{11}{runtime_0}{\runtimeTmp}\rt{\runtimeTmp}
        &\DTLfetchsave{db-fm}{table_index}{12}{runtime_0}{\runtimeTmp}\rt{\runtimeTmp}\\
        \cmidrule(lr){2-7}
        & \multirow{2}{*}{FM$^{+}$} & N
        &\DTLfetch{db-fm}{table_index}{13}{max_constraint}
        &\DTLfetch{db-fm}{table_index}{14}{max_constraint}
        &\DTLfetch{db-fm}{table_index}{15}{max_constraint}
        &\DTLfetch{db-fm}{table_index}{16}{max_constraint}\\
         \cmidrule(lr){3-7}
        & & {$\comptime$}
        &\DTLfetchsave{db-fm}{table_index}{13}{runtime_0}{\runtimeTmp}\rt{\runtimeTmp}
        &\DTLfetchsave{db-fm}{table_index}{14}{runtime_0}{\runtimeTmp}\rt{\runtimeTmp}
        &\DTLfetchsave{db-fm}{table_index}{15}{runtime_0}{\runtimeTmp}\rt{\runtimeTmp}
        &\DTLfetchsave{db-fm}{table_index}{16}{runtime_0}{\runtimeTmp}\rt{\runtimeTmp}\\
		\bottomrule
	\end{tabularx}
\end{table}
During the Fourier-Motzkin elimination process, collection $C'$ potentially contains redundant constraints, creating even more redundancies in the next steps. To optimize this, we identify and remove redundant constraints after each variable elimination. This redundancy check is performed by \hypro \cite{schupp2017hypro}, utilizing the \textsc{GNU Linear Programming Kit} (\textsc{GLPK}) to solve linear programming (LP) problems for each constraint. For each constraint we determine whether it is redundant using the LP solver from GLPK by comparing the state set with and without the constraint. %

\subsubsection{Results} 
Table \ref{tab:both_fm_full} compares results of a maximum reachability analysis~\cite{delicaris2024Journal} with the default Fourier-Motzkin elimination (FM)  and  with Fourier-Motzkin with redundancy checks (FM\(^{+}\)). 
Results indicate that FM\(^{+}\) is slower for most model variants considered but is effective in mitigating constraint explosion.
For instance, the \cebike case study with one cycle resulted in \hpoly{s} with a maximum of 14,457 intermediate constraints and required about 50 minutes for FM, whereas FM\(^{+}\) completed in about 3 minutes with a maximum of 130 intermediate constraints. Constraint explosion in \cebike is probably due to location \textit{high} where two continuous variables have rectangular flow. 

Although redundancy checks can be costly and offer no benefit when no constraint explosion occurs, they greatly improve robustness by preventing constraint blow-up. Therefore, all subsequent computations use FM\(^{+}\).

\subsection{Optionality of Backward Refinement}\label{subsec:backwards_refinement}
As discussed in Section~\ref{subsec:reachability_analysis}, previous approaches use backward reachability~\cite{delicaris2023maximizing}. While our initial focus on scheduler synthesis required it, when only optimal probabilities are of interest, backward computations can be avoided through sufficient bookkeeping of the random variables’ history during forward analysis. Our main theoretical contribution shows that maximum reachability probabilities can be determined using only forward reachability analysis, provided the schedulers themselves are not required.

Let $\RACU$ be an unrolled RAC with goal states $\Goal$. Further, let $\reachtree$ be the $(\tmax,\jumpmax)$-bounded reach tree for $\RACU$ from $\ell$ to $\Goal$, and let $\Goalvali = \symb_i \cap \Goalval{}$ for each $i \in \pathindices$ (see Section~\ref{subsec:reachability_analysis}). 
For every $\Goalvali$, we derive  $\polyfwdi$ by extracting the bounds on the stochastic dimensions, as follows.

\begin{definition}
For each $i \in \pathindices$, we define $\polyfwdi=\{\valRandom\,|\,(\ell,\valCont,\valRandom,\sample)\in\Goalvali\}\subseteq\Reals^{\dimRandomU}$ as $\Goalvali$ projected onto the stochastic dimensions. We define $\polyfwd$ as the union $\polyfwd = \bigcup_{i \in \pathindices} \polyfwdi$.
\end{definition}

Note that the extraction of $\polyfwdi$ necessitates a postprocessing step that accounts for both the lifting procedure (as described in~\cite{delicaris2023maximizing}) and potential race conditions among random clocks.
Lemma \ref{lemma:union} states that $\polyfwd$ encodes $\Prophecies{}^\scheduler$, i.e., the set of all $\jumpmax$-bounded prophecies $\prophecy$ for which $\scheduler_{\prophecy}$ reaches $\Goal$ in $\RACU$ before $\tmax$. To formally specify this correspondence for $\LabRandom=\{\clockr_1,\ldots,\clockr_{\dimRandomU}\}$, let $\trans(\prophecy)=(\prophecy(\clockr_1,0),\ldots,\prophecy(\clockr_{\dimRandomU},0))$ be the sequence of all first sample values for the random clocks under the given prophecy, and let $\trans(\Prophecies{}')=\{\trans(\prophecy)\,|\,\prophecy\in\Prophecies{}'\}$ for a set $\Prophecies{}'$ of prophecies.
\begin{lemma}\label{lemma:union}
Let $\RACU$ with  set of states $\States{\RACU}$ be an unrolled RAC with $\dimRandomU$ random clocks. Assume $\Goal \subseteq \States{\RACU}$, $\tmax \in \Realsposzero$ and $\jumpmax\in\Naturals$. As in Definition \ref{def:probability}, for each scheduler $\scheduler\in\SchedulersProphetic{\RACU}$, let $\Prophecies{}^\scheduler$ be the set of all $\jumpmax$-bounded prophecies $\prophecy$ for which $\scheduler_{\prophecy}$ reaches $\Goal$ in $\RACU$ within bounds $(\tmax,\jumpmax)$. Then, for all maximal schedulers $\scheduler \in \SchedulersProphetic{\RACU}^{\textsl{max}}(\Goal, \tmax, \textsl{jmp})$ it holds \(\trans(\Prophecies{}^\scheduler)=\polyfwd\).

\end{lemma}

The maximal reachability probabilities can be obtained by integrating over the state set \(\polyfwd\). By inserting Lemma \ref{lemma:union} into Definition \ref{def:probability}, we derive the expression for the maximal probability as:
\begin{equation}
\label{eq:maxprob}
p_{\RACU}^\textsl{max}(\Goal, \tmax, \jumpmax) = \int_{\polyfwd} G(\valCont) \, d\valCont.
\end{equation}

The proof of Lemma \ref{lemma:union}, provided in \ifarxiv{Appendix \ref{sec:appendix}}\else{\cite{stuebbe2025arxiv}~}\fi,  follows the approach  in~\cite{delicaris2024Journal}. %
Our contribution shows that forward reachability state sets $\polyfwd$ are sufficient for computing maximal reachability probabilities.

\begin{table}[t]
	\centering

	\centering
	\scriptsize
  \captionof{table}{Computation times for maximum reachability analysis of $\polyfb$ and $\polyfwd$ under different cycle and charging type combinations in the \ccar and \cebike case studies.}
	\label{tab:both_br}
	\newcolumntype{Y}{>{\centering\arraybackslash}X}
         \newcolumntype{P}[1]{>{\centering\arraybackslash}p{#1}}
	\renewcommand{\arraystretch}{1}
\begin{tabularx}{\linewidth}{p{0.265cm}p{0.29cm}YYYYY}
		\toprule
        \ & \ &\multicolumn{3}{c}{\ccar}&\multicolumn{2}{c}{\cebike}\\
        \cmidrule(lr){3-5}\cmidrule(lr){6-7}
        \multicolumn{2}{c}{cycles}& $0$  & $1$  & $2$  & $0$  & $1$ \\
    \multicolumn{2}{c}{type}&\texttt{ABC}&\texttt{ABC}&\texttt{A}&&\\
		\midrule
		\multirow{2}{*}{\rotatebox{90}{rect.}}
		& \multirow{1}{*}{\shortstack[l]{$\polyfb$}}
& \DTLfetchsave{db-backward}{table_index}{1}{runtime_0}{\runtime}\rt{\runtime}
	& \DTLfetchsave{db-backward}{table_index}{2}{runtime_0}{\runtime}\rt{\runtime}
	& \DTLfetchsave{db-backward}{table_index}{3}{runtime_0}{\runtime}\rt{\runtime}
	& \DTLfetchsave{db-backward}{table_index}{4}{runtime_0}{\runtime}\rt{\runtime}
	& \DTLfetchsave{db-backward}{table_index}{5}{runtime_0}{\runtime}\rt{\runtime} \\
	\cmidrule{3-7} 
	& \multirow{1}{*}{\shortstack[l]{$\polyfwd$}}
	& \DTLfetchsave{db-backward}{table_index}{6}{runtime_0}{\runtime}\rt{\runtime}
	& \DTLfetchsave{db-backward}{table_index}{7}{runtime_0}{\runtime}\rt{\runtime}
	& \DTLfetchsave{db-backward}{table_index}{8}{runtime_0}{\runtime}\rt{\runtime}
	& \DTLfetchsave{db-backward}{table_index}{9}{runtime_0}{\runtime}\rt{\runtime}
	& \DTLfetchsave{db-backward}{table_index}{10}{runtime_0}{\runtime}\rt{\runtime} \\
	\midrule
	\multirow{2}{*}{\rotatebox{90}{sing.}}
	& \multirow{1}{*}{\shortstack[l]{$\polyfb$}}
	& \DTLfetchsave{db-backward}{table_index}{11}{runtime_0}{\runtime}\rt{\runtime}
	& \DTLfetchsave{db-backward}{table_index}{12}{runtime_0}{\runtime}\rt{\runtime}
	& \DTLfetchsave{db-backward}{table_index}{13}{runtime_0}{\runtime}\rt{\runtime}
	& \DTLfetchsave{db-backward}{table_index}{14}{runtime_0}{\runtime}\rt{\runtime}
	& \DTLfetchsave{db-backward}{table_index}{15}{runtime_0}{\runtime}\rt{\runtime} \\
	\cmidrule{3-7} 
	& \multirow{1}{*}{\shortstack[l]{$\polyfwd$}} 
	& \DTLfetchsave{db-backward}{table_index}{16}{runtime_0}{\runtime}\rt{\runtime}
	& \DTLfetchsave{db-backward}{table_index}{17}{runtime_0}{\runtime}\rt{\runtime}
	& \DTLfetchsave{db-backward}{table_index}{18}{runtime_0}{\runtime}\rt{\runtime}
	& \DTLfetchsave{db-backward}{table_index}{19}{runtime_0}{\runtime}\rt{\runtime}
	& \DTLfetchsave{db-backward}{table_index}{20}{runtime_0}{\runtime}\rt{\runtime} \\
		\bottomrule
	\end{tabularx}
\end{table}

\begin{figure}[h]
	\input{plots/plots}
\end{figure}

\subsubsection{Results}
Table \ref{tab:both_br} compares computation times for maximum reachability analysis using the  approach presented in~\cite{delicaris2024Journal} (denoted as \(\polyfb\)), with the proposed forward reachability approach (denoted as  \(\polyfwd\)). 
Results indicate that the forward approach achieves speedups for both singular and rectangular variants by a factor of  \round{2.23657089319
} to \round{4.40186615187}  for \ccar and of \round{1.87880794702} to \round{6.33484162896} for \cebike. To better illustrate scalability, we have included the model version \ccar with two cycles in variant \texttt{A}.
Figure \ref{fig:plots_time_breakdown}  illustrates the computation time per invoked method for both computations.  Backward computation consists of the refinement and its preprocessing, which  dominates the total runtime in $\polyfb$. 
For validation, we provide the computed reachability probabilities in \ifarxiv{Appendix \ref{sec:appendix}}\else{\cite{stuebbe2025arxiv}~}\fi. \todo{habe die bezeichner der fehler hier rausgenommen, die kennt man ja noch nicht. }
As expected, the computed probabilities for both case studies with $\polyfwd$ and $\polyfb$ match. %
The results of the \ccar case study align with those of~\cite{delicaris2024Journal}, which compared the previous implementation of \realyst with \prohver. For the \cebike case study with one cycle, computations performed by \prohver did not terminate within 20 hours.

\subsection{Adapted Integration Bounds}\label{subsec:dynamic_integration_bound}
\begin{table*}[t]
	\centering
	\scriptsize
	\caption{%
(Adapted) integration domains for random variables in the \ccar case study with rectangular domains, charging type ABC, and 1 cycle, for $t_{\textsl{int}}=100$ using \samples{1e8} integration samples.
}%
	\label{tab:improvements_dynamic_integration_bound}
	\newcolumntype{Y}{>{\centering\arraybackslash}X}
 \newcolumntype{Z}{>{\centering\arraybackslash}p{1cm}}
 \newcolumntype{W}{>{\centering\arraybackslash}p{1.6cm}}
	\renewcommand{\arraystretch}{1.25}
\begin{tabularx}{\linewidth}{p{0.5cm}YYYYYZW}
		\toprule
          \ & \multicolumn{4}{c}{$F_Y$} & Exp & \multirow{2}{*}{{$\comptime$}} & \multirow{2}{*}{$\estat$}\\
          \cmidrule(lr){2-5}\cmidrule(lr){6-6}
          \ & $\mu=1.5,\sigma=2$&$\mu=2,\sigma=0.75$&$\mu=1.5,\sigma=2$&$\mu=2,\sigma=0.75$&$\lambda=1$ &&\\
        \midrule
$\mathbb{I}_{100}$
	&\DTLfetchsave{db-adaptedIntegrationBounds}{table_index}{1}{bounding_box_lower_1}{\lower}\DTLfetchsave{db-adaptedIntegrationBounds}{table_index}{1}{bounding_box_upper_1}{\upper}[\ifthenelse{\equal{\lower}{0}}{0}{\rintervals{\lower}}, \ifthenelse{\equal{\upper}{100}}{100}{\rintervals{\upper}}]		
		&\DTLfetchsave{db-adaptedIntegrationBounds}{table_index}{1}{bounding_box_lower_2}{\lower}\DTLfetchsave{db-adaptedIntegrationBounds}{table_index}{1}{bounding_box_upper_2}{\upper}[\ifthenelse{\equal{\lower}{0}}{0}{\rintervals{\lower}}, \ifthenelse{\equal{\upper}{100}}{100}{\rintervals{\upper}}]		
		&\DTLfetchsave{db-adaptedIntegrationBounds}{table_index}{1}{bounding_box_lower_4}{\lower}\DTLfetchsave{db-adaptedIntegrationBounds}{table_index}{1}{bounding_box_upper_4}{\upper}[\ifthenelse{\equal{\lower}{0}}{0}{\rintervals{\lower}}, \ifthenelse{\equal{\upper}{100}}{100}{\rintervals{\upper}}]		
		&\DTLfetchsave{db-adaptedIntegrationBounds}{table_index}{1}{bounding_box_lower_5}{\lower}\DTLfetchsave{db-adaptedIntegrationBounds}{table_index}{1}{bounding_box_upper_5}{\upper}[\ifthenelse{\equal{\lower}{0}}{0}{\rintervals{\lower}}, \ifthenelse{\equal{\upper}{100}}{100}{\rintervals{\upper}}]		
		&\DTLfetchsave{db-adaptedIntegrationBounds}{table_index}{1}{bounding_box_lower_3}{\lower}\DTLfetchsave{db-adaptedIntegrationBounds}{table_index}{1}{bounding_box_upper_3}{\upper}[\ifthenelse{\equal{\lower}{0}}{0}{\rintervals{\lower}}, \ifthenelse{\equal{\upper}{100}}{100}{\rintervals{\upper}}]
		&\DTLfetchsave{db-adaptedIntegrationBounds}{table_index}{1}{runtime_0}{\runtimetmp}\rtt{2}{\runtimetmp}
		&\DTLfetchsave{db-adaptedIntegrationBounds}{table_index}{1}{error_stat}{\estattmp}\errortabletfont{2}{\estattmp}\\
		          $\mathbb{I}_{\text{100}}^{\text{adap}}$
		&\DTLfetchsave{db-adaptedIntegrationBounds}{table_index}{2}{bounding_box_lower_1}{\lower}\DTLfetchsave{db-adaptedIntegrationBounds}{table_index}{2}{bounding_box_upper_1}{\upper}[\ifthenelse{\equal{\lower}{0}}{0}{\rintervals{\lower}}, \ifthenelse{\equal{\upper}{100}}{100}{\rintervals{\upper}}]		
		&\DTLfetchsave{db-adaptedIntegrationBounds}{table_index}{2}{bounding_box_lower_2}{\lower}\DTLfetchsave{db-adaptedIntegrationBounds}{table_index}{2}{bounding_box_upper_2}{\upper}[\ifthenelse{\equal{\lower}{0}}{0}{\rintervals{\lower}}, \ifthenelse{\equal{\upper}{100}}{100}{\rintervals{\upper}}]		
		&\DTLfetchsave{db-adaptedIntegrationBounds}{table_index}{2}{bounding_box_lower_4}{\lower}\DTLfetchsave{db-adaptedIntegrationBounds}{table_index}{2}{bounding_box_upper_4}{\upper}[\ifthenelse{\equal{\lower}{0}}{0}{\rintervals{\lower}}, \ifthenelse{\equal{\upper}{100}}{100}{\rintervals{\upper}}]		
		&\DTLfetchsave{db-adaptedIntegrationBounds}{table_index}{2}{bounding_box_lower_5}{\lower}\DTLfetchsave{db-adaptedIntegrationBounds}{table_index}{2}{bounding_box_upper_5}{\upper}[\ifthenelse{\equal{\lower}{0}}{0}{\rintervals{\lower}}, \ifthenelse{\equal{\upper}{100}}{100}{\rintervals{\upper}}]		
		&\DTLfetchsave{db-adaptedIntegrationBounds}{table_index}{2}{bounding_box_lower_3}{\lower}\DTLfetchsave{db-adaptedIntegrationBounds}{table_index}{2}{bounding_box_upper_3}{\upper}[\ifthenelse{\equal{\lower}{0}}{0}{\rintervals{\lower}}, \ifthenelse{\equal{\upper}{100}}{100}{\rintervals{\upper}}]
		&\DTLfetchsave{db-adaptedIntegrationBounds}{table_index}{2}{runtime_0}{\runtimetmp}\rtt{2}{\runtimetmp}
		&\DTLfetchsave{db-adaptedIntegrationBounds}{table_index}{2}{error_stat}{\estattmp}\errortabletfont{2}{\estattmp}\\
    \bottomrule
    \end{tabularx}
\end{table*}

 The maximum reachability probability (see Equation~\ref{eq:maxprob}) is obtained by integration over $\polyfwd$, which is the union of $\Goalvali\proj{\dimRandomU}$ over all $i \in \pathindices$. 
Note that for $\Goalvali$, where $\clockr \in \dimRandomU$ did not expire,  $\Goalvali\proj{\clockr}$ needs to equal $\Reals_{\geq 0}$, such that the entire probability mass from the underlying probability distribution $\CDF_{\clockr}$ is considered. %
Since multidimensional integration over unbounded polyhedra is in practice not possible,  we \emph{lift} to a predefined \emph{integration bound} $\tint \geq \tmax$  and not to infinity, as explained in \cite{delicaris2024Journal}. \todo{why is $\tint$ not $T_{int}$?}
This results in bounded polyhedra, which induces the following error
\begin{equation}\label{eq:e_inf}
e_{\infty}= 1 -  \int_{[0,\tint]^{\dimRandomU}}    \prod_{\clockr\in\dimRandomU} \mathds{1}(\clockr) \Distr(\clockr)(\valCont) d\valCont,
\end{equation}
where $\mathds{1}(\clockr) = 1$ if and only if there exists $i \in \pathindices$ such that $\clockr$ did not expire in $\Goalvali$, and thus was lifted along at least one path $i$.
This error is an overapproximation of the actual error, since dependencies between the random variables are not explicitly covered in the error computation.

Clearly, increasing $\tint$ potentially decreases $e_{\infty}$. %
Integration is done statistically with Monte Carlo Vegas~\cite{lepage1978integration},  which introduces an additional statistical error $\estat$,  depending on the number of integration samples.

Monte Carlo Vegas chooses sample values from  each dimension separately, resulting in samples which lie inside the bounding box $bb \subseteq [0, \tint]^{\dimRandomU}$, but not necessarily inside $\polyfwd$. Especially for larger dimensions this leads to many samples being discarded, as they are not contained in $\polyfwd$ and hence to a larger statistical error $\estat$.
Finding a smaller $t_{\textsl{int}}$ per dimension optimizes performance by decreasing the statistical error \(\estat\) for a fixed number of samples, as fewer samples lie outside $\polyfwd$, or achieving the same error with fewer samples, thus reducing computational runtime. %
For simplicity, $t_{\textsl{int}}$ was set to $100$ for every integration dimension in \cite{delicaris2024Journal}. This is improved  by choosing a dedicated upper integration bound for every dimension. We present an algorithm which tightens upper (and potentially lower) integration bounds  in an automated way.

\subsubsection{Tightening the integration bounds} %

For a specific RAC,  tighter integration bounds can often be identified, denoted by the multi-dimensional interval $[L, R]^{\dimRandomU}$, with $L\proj{\clockr}, R\proj{\clockr} \in \Reals$ and $0 \leq L\proj{\clockr} < R\proj{\clockr} \leq t_{\textsl{int}}$ for every $\clockr \in \dimRandomU$. 
The proposed method heuristically reduces the integration bounds without introducing an additional error, by eliminating areas with zero probability mass, limited by IEEE double precision. This  can be applied independently to each stochastic dimension by determining parameters \( L\proj{\clockr}, R\proj{\clockr} \in \mathbb{R} \) for all \( \clockr \in \dimRandomU \), chosen with smaller distance \( R\proj{\clockr}-L\proj{\clockr} \), which still ensure that
$\int_{[L,R]^{\dimRandomU}} G(s) \, ds = \int_{[0,t_{\textsl{int}}]} G(s) \, ds$. %
We address every stochastic dimension  individually and explain how to tighten the bounds $L, R$  for one random clock. Note that we  omit the projection onto $\clockr$  for readability.

For different distribution $\CDF$,  bounds \(L\) and \(R\) are tightened differently. For a folded normal distribution characterized by \(F_Y(\mu, \sigma)\) and a given initial upper bound \(t_{\textsl{int}}\) Algorithm \ref{alg:dyn_int}  computes tighter bounds. %
Initially, \( L \) and \( R \) are symmetrically positioned around the mean \( \mu \), either at a distance of \( \mu \) (l. 4) or \( \tint \) (l. 8) to $\mu$. This placement may introduce a nonzero error, when $R$ was placed left of $\tint$ or $L$ right of $0$ (l. 5 or 9), which is corrected by adjusting the inducing bound, modifying \( R \) (l. 6) or \( L \) (l. 10) as needed. Finally, the algorithm employs binary search to shift both \( L \) and \( R \) closer to \( \mu \) as much as possible while ensuring no error is introduced (l. 11).

Dedicated algorithms can be derived for other distributions. For example  for a uniform distribution with support \(\support(\CDF) \in [a,b]\), the bounds are defined as \(L = \max(0, a)\) and \(R = \min(b, t_{\textsl{int}})\) before the binary search starts. %

\begin{algorithm}[h]
\scriptsize
\caption{Adapting the integration bounds in one dimension for $F_Y(\mu,\sigma)$}\label{alg:dyn_int}
\begin{algorithmic}[1]
\State \textbf{Input:} $F_Y(\mu, \sigma), t_{\textsl{int}}$
\State $L \gets 0, R \gets t_{\textsl{int}}$

\If{$2\mu - t_{\textsl{int}} < 0$}
    \State $L \gets 0, R \gets 2\mu$

    \If{$\int_{[R,t_{\textsl{int}}]} F_Y(\mu,\sigma)(s) > 0$}
        \State $L \gets 0, R \gets \text{BinarySearch}(F_Y(\mu, \sigma), L:=R, R:=t_{\textsl{int}})$
    \EndIf
\Else
    \State $L \gets 2\mu - t_{\textsl{int}}, R \gets t_{\textsl{int}}$
    \If{$\int_{[0,L]} F_Y(\mu,\sigma)(s) > 0$}
        \State $L \gets \text{BinarySearch}(F_Y(\mu, \sigma), L:=0, R:=L), R \gets t_{\textsl{int}}$
    \EndIf
\EndIf

\State $L \gets \text{FindLarger}(L, F_Y(\mu, \sigma), R:=\mu)$
\State $R \gets \text{FindSmaller}(R, F_Y(\mu, \sigma), L:=\mu)$

\State \textbf{Output:} $[L, R]$
\end{algorithmic}
\end{algorithm}

\subsubsection{Results}

Table \ref{tab:improvements_dynamic_integration_bound} compares the integration domains for the rectangular version of \ccar with charging type \texttt{ABC} and 1 cycle for $\tint=100$ with and without the optimization of the integration bounds. The volume without optimization equals \(\rscientific{1.66667e+09}\), which is approximately 49 times larger than the volume corresponding to the adapted integration bounds. We see that the reduction in volume accelerates computation and improves statistical accuracy. %
Since the chosen method reduces computation cost per sample within the polytope but incurs an initial setup cost, the computation times for \samples{1e7} were still similar (about \rtt{0}{400}) between the optimized and non-optimized variants. Therefore, the number of integration samples was increased to \samples{1e8} to demonstrate that the break-even point is quickly surpassed when targeting a reasonable $\estat$.

\section{Conclusion and Future Work}\label{sec:conclusion}
We prove that forward reachability analysis is able to compute maximum reachability probabilities for RAC,  and show that this results in significantly smaller computation times.  
Further, Fourier-Motzkin quantifier elimination was enhanced with checks for redundant constraints  (FM\(^{+}\)), which is able to  prevent constraint explosion in one model variant. 
Additionally, we present an algorithm which computes tighter integration bounds, resulting in a significantly reduced integration volume and improved efficiency. 
Overall, the presented improvements increase the feasibility of the analysis of rectangular  automata with stochastic delays, paving the way towards automated worst-case analysis of safety-critical systems.

Future work will continue to investigate improvements for the reachability analysis of stochastic hybrid models, e.g. by re-using dimensions of the state-space for random variables, to make the flowpipe analysis more efficient.

\ifarxiv
\cleardoublepage
\begin{appendix}
\renewcommand{\thesection}{Appendix \Alph{section}}
\section{}%
\label{sec:appendix}
\label{sec:appendix_definitions}
The additional material contained in all appendices will be made available
on arXiv upon acceptance of the paper.

\def\ROUTEDISTANCE{20}
\def\MAXCAPACITY{60}
\def\SERVICETIME{15}

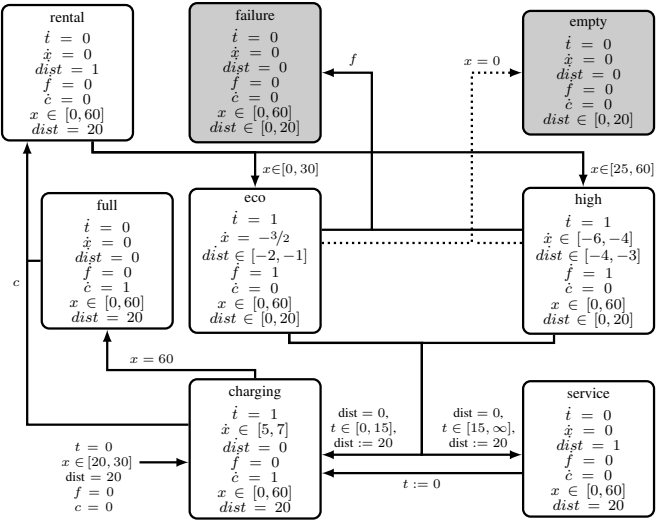
\begin{figure}[htb]
\def\singular{0}
\def\advancedFlowRepresentation{1}
\def\showInvariants{1}
 \scalebox{0.715}{%
\begin{tikzpicture}[
	n/.style={draw, text width = 1.3cm, minimum height = 1.7cm, align = center, font= \footnotesize, rounded corners, very thick, execute at begin node=\setlength{\baselineskip}{8pt}%
	},
	nsmall/.style={draw, text width = 2.2cm, minimum height = 1cm, align = center, font= \footnotesize, rounded corners, very thick, execute at begin node=\setlength{\baselineskip}{8pt}%
	},
	charging/.style={minimum height = 1.3cm
	},
	en/.style={draw=none, minimum height=0cm, font = \scriptsize, align = center},%
	c/.style={draw, fill, black, circle, inner sep=0, outer sep=0, minimum size=1mm},
	l/.style={anchor=west, inner sep=0, font=\footnotesize},
 ]

	\node[nsmall,charging](eco) at (-2,0) {
		$\text{eco}$	\\
		\vspace{0.125cm} 	
		$\dot{t}=1$\\
		\if\singular1
		$\dot{x}=-1$\\
		$\dot{dist}=-1$\\
		\else
		\vspace{0.05cm}$\dot{x}=-\nicefrac{3}{2}$\\
		$\dot{dist}\in[-2,-1]$\\
		\fi
\if\advancedFlowRepresentation1
$\dot{f}=1$\\
$\dot{c}=0$\\
\else
\fi
	\if\showInvariants1
		$x\in[0,\MAXCAPACITY]$
		$dist\in[0,\ROUTEDISTANCE]$\\
		\else\fi
	};

	\node[nsmall,charging](high) at (4.2,0) {
	$\text{high}$	\\
	\vspace{0.125cm} 	
	$\dot{t}=1$\\
		\if\singular1
$\dot{x}=-5$\\
$\dot{dist}=-3$\\
\else
$\dot{x}\in[-6,-4]$\\
$\dot{dist}\in[-4,-3]$\\
\fi
\if\advancedFlowRepresentation1
$\dot{f}=1$\\
$\dot{c}=0$\\
\else
\fi
	\if\showInvariants1
	$x\in[0,\MAXCAPACITY]$
	$dist\in[0,\ROUTEDISTANCE]$\\
	\else\fi
};

	\node[nsmall,charging](service) at (4.2,-3.5) {
	$\text{service}$	\\
	\vspace{0.125cm} 	
	$\dot{t}=0$\\
	$\dot{x}=0$\\
	$\dot{dist}=1$\\
\if\advancedFlowRepresentation1
$\dot{f}=0$\\
$\dot{c}=0$\\
\else
\fi
	\if\showInvariants1
	$x\in[0,\MAXCAPACITY]$
	$dist=\ROUTEDISTANCE$\\
	\else\fi
};

	\node[nsmall,charging](charging) at (-2,-3.5) {
	$\text{charging}$	\\
	\vspace{0.125cm} 	
	$\dot{t}=1$\\
			\if\singular1
	$\dot{x}=5$\\
	\else
	$\dot{x}\in[5,7]$\\
	\fi
	$\dot{dist}=0$\\
\if\advancedFlowRepresentation1
$\dot{f}=0$\\
$\dot{c}=1$\\
\else
\fi
	\if\showInvariants1
$x\in[0,\MAXCAPACITY]$
$dist=\ROUTEDISTANCE$\\
\else\fi
};

	\node[nsmall,charging, fill=maincolor!20!white](empty) at (4.2,3.5) {
	$\text{empty}$	\\
	\vspace{0.125cm} 	
	$\dot{t}=0$\\
	$\dot{x}=0$\\
	$\dot{dist}=0$\\
\if\advancedFlowRepresentation1
$\dot{f}=0$\\
$\dot{c}=0$\\
\else
\fi
\if\showInvariants1
	$dist\in[0,\ROUTEDISTANCE]$
\else\fi
};
	
	\node[nsmall,charging, fill=maincolor!20!white](failure) at (-2,3.5) {
	$\text{failure}$	\\
	\vspace{0.125cm} 	
$\dot{t}=0$\\
$\dot{x}=0$\\
$\dot{dist}=0$\\
\if\advancedFlowRepresentation1
$\dot{f}=0$\\
$\dot{c}=0$\\
\else
\fi
	\if\showInvariants1
$x\in[0,\MAXCAPACITY]$
$dist\in[0,\ROUTEDISTANCE]$\\
\else\fi
};	

	\node[nsmall,charging](full) at (-4.75,0) {
	$\text{full}$	\\
	\vspace{0.125cm} 	
	$\dot{t}=0$\\
	$\dot{x}=0$\\
	$\dot{dist}=0$\\
	\if\advancedFlowRepresentation1
	$\dot{f}=0$\\
	$\dot{c}=1$\\
	\else
	\fi
	\if\showInvariants1
$x\in[0,\MAXCAPACITY]$
$dist=\ROUTEDISTANCE$\\
	\else
\fi};

	\node[nsmall,charging](start) at ($(-2.25,0)+(failure.180)$) {
	$\text{rental}$	\\
	\vspace{0.125cm} 	
	$\dot{t}=0$\\
	$\dot{x}=0$\\
	$\dot{dist}=1$\\
\if\advancedFlowRepresentation1
$\dot{f}=0$\\
$\dot{c}=0$\\
\else
\fi
	\if\showInvariants1
$x\in[0,\MAXCAPACITY]$
$dist=\ROUTEDISTANCE$\\
\else\fi};

\draw[-latex,  very thick] (eco.295) -- ($(0,-0.15) + (eco.295 |- , |- high.south)$) -- ($(0,-0.15) + ( {$(eco)!0.5!(high)$} |- , |- high.south)$) -- ( {$(eco)!0.5!(high)$} |- 99 , 99 |- service.185) to node[en, above right=0.0cm and -0.7cm] {$\text{dist} = 0$,\\$t \in [\SERVICETIME, \infty]$,\\$\text{dist} := \ROUTEDISTANCE$} (service.185);

\draw[-latex,  very thick] (eco.295) -- ($(0,-0.15) + (eco.295 |- , |- high.south)$) -- ($(0,-0.15) + ( {$(eco)!0.5!(high)$} |- , |- high.south)$) -- ( {$(eco)!0.5!(high)$} |- 99 , 99 |- service.185) to node[en, above left=0.0cm  and -0.6cm] {$\text{dist} = 0$,\\$t \in [0, \SERVICETIME]$,\\$\text{dist} := \ROUTEDISTANCE$} (charging.355);

\draw[very thick] (high.245) -- ($(0,-0.15) + (high.245 |- , |- high.south)$) -- ($(0,-0.15) + ( {$(eco)!0.5!(high)$} |- , |- high.south)$);

\draw[-latex,  very thick] (service.200) to node[en, below=-0.025cm] {$t:=0$} ( charging.340 |- 99 , 99 |- service.200);
	
\draw[-latex,  very thick] (charging.north) -- ($(0,0.15) + (charging.north |- , |- charging.north)$) to node[en, above left=0cm and 0cm] {$x=\MAXCAPACITY$} ($(0,0.15) +( full.270 |- 99 , 99 |- charging.north)$) -- ( full.270 |- 99 , 99 |- full.270);

 \draw[-latex,  very thick] (charging.160) -- ($(-0.25,0) + (full.west |- , |- charging.160)$) to node[en, left] {$c$} ($(-0.25,0) + (full.west |- , |- start.270)$);

\draw[very thick] (full.west) -- ($(-0.25,0) + (full.west |- , |- full.west)$);

\draw[-latex,  very thick] (start.290) -- ($(0,-0.2) + (start.290)$) -- ($(0,-0.2) + (eco.north |- , |- start.290)$) to node[en, right] {$x{\in}[0,30]$} ($(eco.north) + (0,0)$);

\draw[-latex,  very thick] (start.290) -- ($(0,-0.2) + (start.290)$) -- ($(-0.1,-0.2) + (high.north |- , |- start.290)$) to node[en, right] {$x{\in}[25,60]$} ($(high.north) + (-0.1,0)$);

\draw[-latex,  very thick, dotted] (eco.15) -- ($(0,0) + ( {$(eco)!0.65!(high)$} |- , |- eco.15)$) -- ($(0,0) + ( {$(eco)!0.65!(high)$} |- , |- empty.180)$) to node[en, above left=0.0cm and -0.2cm] {$x=0$} (empty.180);
\draw[very thick, dotted] (high.165) -- ($(0,0) + ( {$(eco)!0.66!(high)$} |- , |- eco.15)$);

\draw[-latex,  very thick] (eco.25) -- ($(0,0) + ( {$(eco)!0.35!(high)$} |- , |- eco.25)$) -- ($(0,0) + ( {$(eco)!0.35!(high)$} |- , |- failure.0)$) to node[en, above right=0.0cm and -0.1cm] {$f$} (failure.0);

\draw[very thick] (high.155) -- ($(0,0) + ( {$(eco)!0.35!(high)$} |- , |- eco.25)$);
	
\node[en, text width=1.2cm,execute at begin node=\setlength{\baselineskip}{8pt}, anchor=north] (init) at (-5,-3.25) {$t=0$\\
	$x\in${$\,[20, 30]$}\\
	$\text{dist}=\ROUTEDISTANCE$
    \\$f=0$\\$c=0$
    };
\draw[-latex,  very thick] ($(0.6,-0.25) + ( full.south |- 99 , 99 |- charging.180)$) to node[en, above] {} ($(0,-0.25) + (charging.180)$);
	
\end{tikzpicture}
}
\caption{\cebike model with cycles modeled as RAC.}
\label{fig:ebike_casestudy_full_appendix}
\end{figure}

\begin{definition}[Run of $\RAC$~\cite{delicaris2024Journal}]
A (finite) \emph{run} of $\RAC$ is a finite sequence $\Path=\hastate_0\xrightarrow{a_1}\hastate_1\xrightarrow{a_2}\ldots\hastate_n$, such that for all $i\in\{0,\ldots,n\}$: $\hastate_i=(\ell_i,\valCont_i, \valRandom_i, \sample_i)\in\States{\RAC}$, $\valCont_0\in\Inv(\ell_0)$, $a_{i}\in \Reals_{\geq0}\cup\Edge$, and $\hastate_{i-1}\semanticsarrowrac{a_{i}}\hastate_{i}$ for $i \neq 0$. We require that each time step, i.e., $a_i=t_i\in \mathbb{R}_{\geq0}$  is \emph{maximal}, such that either (i) $t\leq t_i$ for any $\hastate_{i-1}\semanticsarrowrac{t}\hastate_{i}'$, or (ii)  $i<n$ and the time step is followed by a jump $a_{i+1}\in\Edge$, or (iii)  $i=n$ and the time step can be followed by a jump, i.e., $\hastate_n\semanticsarrowrac{e}\hastate_{n+1}$ for some jump $e$ and state $\hastate_{n+1}$. 
We define $ | \Path | = n$, $\last(\Path)=\hastate_n$ and let $\Paths{\RAC}$ be the set of all runs of $\RAC$. 

We call $\Path$ \emph{initial} if $\valCont_0\in\Init(\ell_0)$,  $\valRandom_{0_{\clockr}} = 0$ and $\sample_{0_\clockr} \in \support(\Distr(\clockr))$ for all $\clockr \in \LabRandom$.
Let $\InitPaths{\RAC}$ be the set of all initial runs of $\RAC$.
A state $\hastate=(\ell,\valCont)$ is \emph{reachable} if there exists an initial run $\Path=\hastate_0\xrightarrow{a_1}\hastate_1\xrightarrow{a_2}\ldots\hastate_n$, such that either $\hastate_n=\hastate$ or $n>0$, $\hastate_{n-1}\semanticsarrowrac{t,\textit{rate}}\hastate_n$ and there exists $0<t'<t$ with $\valCont=\valCont_{n-1}+t'\cdot\textit{rate}$.
\end{definition}

\begin{definition}[Scheduler~\cite{delicaris2024Journal}]
 \label{def:SchedulerAppendix}
 \sloppy A \emph{(prophetic history-dependent) scheduler $\scheduler$ for $\RAC$} defines for each prophecy $\prophecy\in\Prophecies{\RAC}$ a function  $\allowbreak\scheduler_{\prophecy} :\allowbreak (\InitPaths{\RAC}\cup\{\epsilon\}) \allowbreak\rightarrow  (\Loc\times\Reals^d)\cup  (\Realsposzero\times \Reals^\dimCont)\cup  (\Edge\times(\Reals\cup\{id\})^{\dimCont})$, such that $\scheduler_{\prophecy}(\epsilon)\in\mathit{InitialChoices}_{\RAC}$ and $\scheduler_{\prophecy}(\Path) \in \JumpChoices_{\RAC}(\last(\Path)) \cup \TimeChoices_{\RAC}(\Path)$ for every initial run  $\Path$ of $\RAC$.
Let $\SchedulersProphetic{\RAC}$ be the set of all schedulers for $\RAC$.
 \end{definition} 

 \begin{definition}[Run induced by scheduler $\scheduler_{\prophecy}$ for RAC $\RAC$~\cite{delicaris2024Journal}]
For a RAC $\RAC$, a scheduler $\scheduler$ for $\RAC$, a prophecy $\prophecy$, and a fixed $n\in\Naturals$, we define \emph{the run of length $n$ induced by $\scheduler_{\prophecy}$ in $\RAC$} recursively to be the unique initial run $\Path(\RAC,\scheduler_{\prophecy},n)\in\InitPaths{\RAC}$ of length $n$ such that:

\begin{itemize}

\item If $n=0$ then $\Path(\RAC,\scheduler_{\prophecy},0)=(\ell,\valCont,\valRandom,\sample)$ with $(\ell,\valCont)=\scheduler_{\prophecy}(\epsilon)$, $\valRandom\proj{r}=0$ and $\sample\proj{r}=\prophecy(r,0)$ for all $r\in\LabRandom$.

\item For $n>0$, let $\Path_{n-1}=\Path(\RAC,\scheduler_{\prophecy},n{-}1)$ with $\last(\Path_{n-1})=\hastate=(\ell,\valCont,\valRandom,\sample)$, and assume $\scheduler_{\prophecy}(\last(\Path_{n-1}))=(a,v)$. 
We set $\Path(\RAC,\scheduler_{\prophecy},n)=\Path_{n-1}\xrightarrow{a}\hastate'$, where $\hastate'=(\ell',\valCont',\valRandom',\sample')$ is the unique state of $\RAC$ with $\last(\Path_{n-1})\semanticsarrowrac{a,v}\hastate'$ and such that either (i) $a\in\Reals_{\geq 0}$ or ($a \in \Edge$ and $\Event(a)= \,\norc$), and $\sample=\sample'$, or (ii) $a\in\Edge$, $\Event(a)=r\in \LabRandom$ and $\sample'\proj{r}=\prophecy(r,k)$, where $k = \big| \bigl\{i\in\{1,\ldots,n-1\}\mid a_i\in\Edge\wedge\Event(a_i)=r\bigr\} \big| + 1$. 
\end{itemize}

For $\tmax\in\Reals_{\geq 0}$, $\jumpmax\in\Naturals$ and $\Goal \subseteq \States{\RAC}$, we say that \emph{$\scheduler_{\prophecy}$ reaches $\Goal$ in $\RAC$ within bounds $(\tmax,\jumpmax)$} iff there is $n\in\Naturals$, $\Path=\Path(\RAC,\scheduler_{\prophecy},n)=\hastate_0\xrightarrow{a_1}\hastate_1\xrightarrow{a_2}\ldots\hastate_n$ such that (i) $\big|\bigl\{i\in\Naturals_{\leq n}\mid a_i\in\Edge\wedge\Event(a_i)=r\bigr\}\big| \leq\jumpmax$ for each $r\in\LabRandom$, and (ii) either $\hastate_n\in\Goal$ and $\dur(\Path)\leq\tmax$, or $n>0$, $a_n=t\in\Reals_{\geq 0}$, and there exists $0\leq t'<t$ and $\hastate_n'\in\Goal$ with $\hastate_{n-1}\semanticsarrowrac{t'}\hastate_n'$ and $\dur(\hastate_0\xrightarrow{a_1}\hastate_1\xrightarrow{a_2}\ldots\hastate_{n-1}\xrightarrow{t'}\hastate_n')\leq\tmax$.

\end{definition}

\begin{figure*}[htb]
    \centering
\[
\begin{array}{c}
\\\\
\inference[\texttt{Rule}\textsubscript{$\mathit{Flow}$}]
{
  (\ell,\valCont, \valRandom, \sample) \in \States{\RAC} \quad
  t\in\Realsposzero \quad
  \textit{rate} \in  \FlowCont(\ell) \\
  \valCont'=\valCont+t\cdot \textit{rate} \in \Inv(\ell)\quad
  \valRandom' = \valRandom + t\cdot \FlowRandom{\RAC}(\ell) \leq \sample 
}{
    (\ell,\valCont, \valRandom, \sample) \semanticsarrowrac{t} (\ell,\valCont',\valRandom', \sample)
}
\\[0.5cm]
\inference[\texttt{Rule}\textsubscript{$\mathit{Jump}_{\tinyN}$}]
{
  (\ell,\valCont, \valRandom, \sample) \in \States{\RAC} \quad
  e=(\ell,\guard  , \reset , \ell') \in \Edge \quad
  \Event(e)=\norc \\
  \valCont \in \guard \quad
  \valCont'\in\reset(\valCont)
}{
    (\ell,\valCont, \valRandom, \sample) \semanticsarrowrac{e} (\ell',\valCont',\valRandom, \sample)
}
\\[0.5cm]
\inference[\texttt{Rule}\textsubscript{$\mathit{Jump}_{\tinyS}$}]
{
  (\ell,\valCont, \valRandom, \sample) \in \States{\RAC} \quad
  e=(\ell,\guard  , \reset , \ell') \in \Edge \quad
  \Event(e)=r \in \LabRandom \\
  \valCont'\in\reset(\valCont) \quad
  \valRandom\proj{\clockr} = \sample\proj{\clockr} \quad
}{
    (\ell,\valCont, \valRandom, \sample) \semanticsarrowrac{e} (\ell',\valCont',\valRandom, \sample)
}
\end{array}
\]
\vspace*{-2ex}
    \caption{Operational semantics for unrolled RAC $\RACU= (\RA,\LabRandom,\allowbreak\Distr,\Event)$ with $\RA=(\Loc, \allowbreak\VarCont, \allowbreak\Inv,  \allowbreak\Init,\allowbreak\FlowCont,\Edge)$.
    }
    \label{fig:operationalsemanticsRACUnrolledAppendix}
\end{figure*}

\section{}%
\label{sec:appendix_casestudy}

\subsubsection{Case study}
To ensure completeness, we present the full RAC, which models the \cebike system with random clocks \(c\) and \(f\) in Figure \ref{fig:ebike_casestudy_full_appendix}. This approach aligns with the feasibility study and incorporates all variables and flow dynamics. As a result of the unrolling process, the case study includes 7 locations and 5 dimensions for 0 cycles, while for 1 cycle, it increases to 20 locations and 7 dimensions.

\section{}%
\label{sec:appendix_proof}

\begin{IEEEproof}
The proof of Lemma \ref{lemma:union} closely follows the approach outlined in~\cite{delicaris2024Journal}. While the original lemma in that work relies on combined forward and backward reachability state sets $\polyfb$, our contribution shows that forward reachability state sets $\polyfwd$ are sufficient for computing maximal reachability probabilities.

``$\subseteq$'': Assume $\valRandom\in \trans({\Prophecies{}^\scheduler})$. By definition, under the prophecy $\valRandom$, a maximal scheduler can reach the goal within the given bounds. Thus there is a run of $\RACU$ from an initial state to a goal state, where all random events occur at the predicted times. Due to the \changed{definition of $\RACU$}, these times are recorded in the final valuation of the run.

Due to the soundness of the reachability analysis, this path is represented in the reach tree. That means, there is a $\symb_i = (\ell, \mathcal{P}) \in\Goalreached$ with $\valRandom \in \mathcal{P}$. \changed{Notably, $\symb_i$ is derived solely from the forward reachability analysis.} Consequently, $\valRandom\in\changed{\polyfwdi}$, and thus $\valRandom\in\changed{\polyfwd}$.

``$\supseteq$'': Assume $\valRandom\in\changed{\polyfwd}$, i.e. $\valRandom\in\changed{\polyfwdi}$ for some $i$. That means, the path to node $i$ in the reach tree represents a run of $\RACU$ which reaches the goal via random events at the times as fixed in $\valRandom$. Thus a maximal scheduler can reach the goal under these expiration times, i.e., under a corresponding prophecy.

\end{IEEEproof}

\section{}%
\label{sec:appendix_todo}

Figure \ref{fig:operationalsemanticsRACUnrolledAppendix} illustrates the operational semantics for an unrolled RAC $\RACU= (\RA,\LabRandom,\allowbreak\Distr,\Event)$ with $\RA=(\Loc, \allowbreak\VarCont, \allowbreak\Inv,  \allowbreak\Init,\allowbreak\FlowCont,\Edge)$ $\RACU$, in which random clocks are never reset.

\section{}%
\label{sec:appendix_tables}

\begin{table*}[htb]
	\centering
	\scriptsize
	\caption{%
 Maximum reachability probabilities computed via state sets $\polyfb$ and $\polyfwd$. Results were obtained with \realyst for variants of the \ccar and \cebike case study.
 }%
	\label{tab:both_br_full}
	\newcolumntype{Y}{>{\centering\arraybackslash}X}
	\renewcommand{\arraystretch}{1}
\begin{tabularx}{\linewidth}{p{0.5cm}p{0.7cm}p{0.7cm}YYYYY}
		\toprule
        &&&\multicolumn{3}{c}{\ccar}&\multicolumn{2}{c}{\cebike}\\
        \cmidrule(lr){4-6}\cmidrule(lr){7-8}
		\multicolumn{3}{c}{\# cycles}& $0$ & $1$& $2$& $0$& $1$\\
    &&&\texttt{ABC}&\texttt{ABC}&\texttt{A}&&\\
		\midrule
		\multirow{8}{*}{\rotatebox{90}{rectangular}}
		& \multirow{4}{*}{\shortstack[l]{$\polyfb$}}
& $p_\textsl{max}$ 
& \DTLfetchsave{db-backward}{table_index}{1}{probability}{\probability}\probab{\probability}
& \DTLfetchsave{db-backward}{table_index}{2}{probability}{\probability}\probab{\probability}
& \DTLfetchsave{db-backward}{table_index}{3}{probability}{\probability}\probab{\probability}
& \DTLfetchsave{db-backward}{table_index}{4}{probability}{\probability}\probab{\probability}
& \DTLfetchsave{db-backward}{table_index}{5}{probability}{\probability}\probab{\probability}\\
& & $\estat$ 
& \DTLfetchsave{db-backward}{table_index}{1}{error_stat}{\errorstat}\errortable{\errorstat}
& \DTLfetchsave{db-backward}{table_index}{2}{error_stat}{\errorstat}\errortable{\errorstat}
& \DTLfetchsave{db-backward}{table_index}{3}{error_stat}{\errorstat}\errortable{\errorstat}
& \DTLfetchsave{db-backward}{table_index}{4}{error_stat}{\errorstat}\errortable{\errorstat}
& \DTLfetchsave{db-backward}{table_index}{5}{error_stat}{\errorstat}\errortable{\errorstat}\\
& & $e_{\infty}$  
& \DTLfetchsave{db-backward}{table_index}{1}{error_infty}{\errorinfty}\errortable{\errorinfty}
& \DTLfetchsave{db-backward}{table_index}{2}{error_infty}{\errorinfty}\errortable{\errorinfty}
& \DTLfetchsave{db-backward}{table_index}{3}{error_infty}{\errorinfty}\errortable{\errorinfty}
& \DTLfetchsave{db-backward}{table_index}{4}{error_infty}{\errorinfty}\errortable{\errorinfty}
& \DTLfetchsave{db-backward}{table_index}{5}{error_infty}{\errorinfty}\errortable{\errorinfty}\\
& & {$\comptime$} 
& \DTLfetchsave{db-backward}{table_index}{1}{runtime_0}{\runtime}\rt{\runtime}
& \DTLfetchsave{db-backward}{table_index}{2}{runtime_0}{\runtime}\rt{\runtime}
& \DTLfetchsave{db-backward}{table_index}{3}{runtime_0}{\runtime}\rt{\runtime}
& \DTLfetchsave{db-backward}{table_index}{4}{runtime_0}{\runtime}\rt{\runtime}
& \DTLfetchsave{db-backward}{table_index}{5}{runtime_0}{\runtime}\rt{\runtime}\\
		\cmidrule{2-8} 
		& \multirow{4}{*}{\shortstack[l]{$\polyfwd$}}
  & $p_\textsl{max}$  
& \DTLfetchsave{db-backward}{table_index}{6}{probability}{\probability}\probab{\probability}
& \DTLfetchsave{db-backward}{table_index}{7}{probability}{\probability}\probab{\probability}
& \DTLfetchsave{db-backward}{table_index}{8}{probability}{\probability}\probab{\probability}
& \DTLfetchsave{db-backward}{table_index}{9}{probability}{\probability}\probab{\probability}
& \DTLfetchsave{db-backward}{table_index}{10}{probability}{\probability}\probab{\probability}\\
& & $\estat$ 
& \DTLfetchsave{db-backward}{table_index}{6}{error_stat}{\errorstat}\errortable{\errorstat}
& \DTLfetchsave{db-backward}{table_index}{7}{error_stat}{\errorstat}\errortable{\errorstat}
& \DTLfetchsave{db-backward}{table_index}{8}{error_stat}{\errorstat}\errortable{\errorstat}
& \DTLfetchsave{db-backward}{table_index}{9}{error_stat}{\errorstat}\errortable{\errorstat}
& \DTLfetchsave{db-backward}{table_index}{10}{error_stat}{\errorstat}\errortable{\errorstat}\\
& & $e_{\infty}$ 
& \DTLfetchsave{db-backward}{table_index}{6}{error_infty}{\errorinfty}\errortable{\errorinfty}
& \DTLfetchsave{db-backward}{table_index}{7}{error_infty}{\errorinfty}\errortable{\errorinfty}
& \DTLfetchsave{db-backward}{table_index}{8}{error_infty}{\errorinfty}\errortable{\errorinfty}
& \DTLfetchsave{db-backward}{table_index}{9}{error_infty}{\errorinfty}\errortable{\errorinfty}
& \DTLfetchsave{db-backward}{table_index}{10}{error_infty}{\errorinfty}\errortable{\errorinfty}\\
& & {$\comptime$} 
& \DTLfetchsave{db-backward}{table_index}{6}{runtime_0}{\runtime}\rt{\runtime}
& \DTLfetchsave{db-backward}{table_index}{7}{runtime_0}{\runtime}\rt{\runtime}
& \DTLfetchsave{db-backward}{table_index}{8}{runtime_0}{\runtime}\rt{\runtime}
& \DTLfetchsave{db-backward}{table_index}{9}{runtime_0}{\runtime}\rt{\runtime}
& \DTLfetchsave{db-backward}{table_index}{10}{runtime_0}{\runtime}\rt{\runtime}\\
		\midrule
		\multirow{8}{*}{\rotatebox{90}{singular}}
		& \multirow{4}{*}{\shortstack[l]{$\polyfb$}}
& $p_\textsl{max}$ 
& \DTLfetchsave{db-backward}{table_index}{11}{probability}{\probability}\probab{\probability}
& \DTLfetchsave{db-backward}{table_index}{12}{probability}{\probability}\probab{\probability}
& \DTLfetchsave{db-backward}{table_index}{13}{probability}{\probability}\probab{\probability}
& \DTLfetchsave{db-backward}{table_index}{14}{probability}{\probability}\probab{\probability}
& \DTLfetchsave{db-backward}{table_index}{15}{probability}{\probability}\probab{\probability}\\
& & $\estat$ 
& \DTLfetchsave{db-backward}{table_index}{11}{error_stat}{\errorstat}\errortable{\errorstat}
& \DTLfetchsave{db-backward}{table_index}{12}{error_stat}{\errorstat}\errortable{\errorstat}
& \DTLfetchsave{db-backward}{table_index}{13}{error_stat}{\errorstat}\errortable{\errorstat}
& \DTLfetchsave{db-backward}{table_index}{14}{error_stat}{\errorstat}\errortable{\errorstat}
& \DTLfetchsave{db-backward}{table_index}{15}{error_stat}{\errorstat}\errortable{\errorstat}\\
& & $e_{\infty}$ 
& \DTLfetchsave{db-backward}{table_index}{11}{error_infty}{\errorinfty}\errortable{\errorinfty}
& \DTLfetchsave{db-backward}{table_index}{12}{error_infty}{\errorinfty}\errortable{\errorinfty}
& \DTLfetchsave{db-backward}{table_index}{13}{error_infty}{\errorinfty}\errortable{\errorinfty}
& \DTLfetchsave{db-backward}{table_index}{14}{error_infty}{\errorinfty}\errortable{\errorinfty}
& \DTLfetchsave{db-backward}{table_index}{15}{error_infty}{\errorinfty}\errortable{\errorinfty}\\
& & {$\comptime$} 
& \DTLfetchsave{db-backward}{table_index}{11}{runtime_0}{\runtime}\rt{\runtime}
& \DTLfetchsave{db-backward}{table_index}{12}{runtime_0}{\runtime}\rt{\runtime}
& \DTLfetchsave{db-backward}{table_index}{13}{runtime_0}{\runtime}\rt{\runtime}
& \DTLfetchsave{db-backward}{table_index}{14}{runtime_0}{\runtime}\rt{\runtime}
& \DTLfetchsave{db-backward}{table_index}{15}{runtime_0}{\runtime}\rt{\runtime}\\
		\cmidrule{2-8} 
		& \multirow{4}{*}{\shortstack[l]{$\polyfwd$}} 
& $p_\textsl{max}$ 
& \DTLfetchsave{db-backward}{table_index}{16}{probability}{\probability}\probab{\probability}
& \DTLfetchsave{db-backward}{table_index}{17}{probability}{\probability}\probab{\probability}
& \DTLfetchsave{db-backward}{table_index}{18}{probability}{\probability}\probab{\probability}
& \DTLfetchsave{db-backward}{table_index}{19}{probability}{\probability}\probab{\probability}
& \DTLfetchsave{db-backward}{table_index}{20}{probability}{\probability}\probab{\probability}\\
& & $\estat$ 
& \DTLfetchsave{db-backward}{table_index}{16}{error_stat}{\errorstat}\errortable{\errorstat}
& \DTLfetchsave{db-backward}{table_index}{17}{error_stat}{\errorstat}\errortable{\errorstat}
& \DTLfetchsave{db-backward}{table_index}{18}{error_stat}{\errorstat}\errortable{\errorstat}
& \DTLfetchsave{db-backward}{table_index}{19}{error_stat}{\errorstat}\errortable{\errorstat}
& \DTLfetchsave{db-backward}{table_index}{20}{error_stat}{\errorstat}\errortable{\errorstat}\\
& & $e_{\infty}$ 
& \DTLfetchsave{db-backward}{table_index}{16}{error_infty}{\errorinfty}\errortable{\errorinfty}
& \DTLfetchsave{db-backward}{table_index}{17}{error_infty}{\errorinfty}\errortable{\errorinfty}
& \DTLfetchsave{db-backward}{table_index}{18}{error_infty}{\errorinfty}\errortable{\errorinfty}
& \DTLfetchsave{db-backward}{table_index}{19}{error_infty}{\errorinfty}\errortable{\errorinfty}
& \DTLfetchsave{db-backward}{table_index}{20}{error_infty}{\errorinfty}\errortable{\errorinfty}\\
& & {$\comptime$} 
& \DTLfetchsave{db-backward}{table_index}{16}{runtime_0}{\runtime}\rt{\runtime}
& \DTLfetchsave{db-backward}{table_index}{17}{runtime_0}{\runtime}\rt{\runtime}
& \DTLfetchsave{db-backward}{table_index}{18}{runtime_0}{\runtime}\rt{\runtime}
& \DTLfetchsave{db-backward}{table_index}{19}{runtime_0}{\runtime}\rt{\runtime}
& \DTLfetchsave{db-backward}{table_index}{20}{runtime_0}{\runtime}\rt{\runtime}\\
		\bottomrule
	\end{tabularx}
\end{table*}

Table \ref{tab:both_br_full} provides the results of the maximum reachability analysis, computed using both the forward (\(\polyfb\)) and forward-backward (\(\polyfwd\)) approaches, for various versions of the \ccar and \cebike case studies. For reproducibility, the integration procedure for all results in this paper was executed with the default seed value of \gsl, which is 0.

\section{}%
\label{sec:appendix_runningexample}

\subsubsection{Running example}
Across all paths, the system starts with a charge level \( x \in [1, 2] \) and charges at a rate between \(4\) and \(6\), 
before discharging at a rate from the interval \([6, 9]\). (i) Path 1: \emph{charge $\rightarrow$ drive $\rightarrow$ charge $\rightarrow$ drive $\rightarrow$ empty.} On this path, the system must not deplete its charge during the first \emph{drive} phase. Hence, the charge level must reach at least \(6\) before entering \emph{drive} for the first time. Starting at \( x = 2 \) and charging at the maximal rate of \(6\) yields 
\( x = 2 + 6 \cdot t = 6 \) after \( t = \tfrac{2}{3} \) time units, which defines the lower bound on \( c_0 \).

Conversely, the system cannot become fully charged on this path. Thus, the first \emph{charge} phase must end before \( x = 10 \). Starting from \( x = 1 \) and charging at rate \(4\), this corresponds to a maximum stay time of \(2.25\) time units in the first \emph{charge} location, which defines the upper bound on \(c_0\).

In the second cycle, \(x\) must not exceed \(9\) or the system would be unable to reach the \emph{empty} state. Starting again from \(x = 0\), this limits the stay time in \emph{charge} to at most \(2.25\) time units. There is also a dependency between both charging phases given by $c_0 + c_1 \leq 4.25$. If the system fully utilizes the first \(2.25\) time units (reaching \(x = 10\)), it returns with \(x = 1\) after \emph{drive} and can then remain only up to \(2\) time units in \emph{charge} to stay below \(x = 9\). Conversely, if the first \emph{charge} phase lasts up to \(2\) time units, the system enters the next \emph{charge} phase with \(x = 9\), allowing up to \(2.25\) time units before returning to \emph{drive}. Hence,
\[c_0 \in \bigl[\tfrac{2}{3}, 2.25\bigr], \quad
  c_1 \in [0, 2.25], \quad
  c_0 + c_1 \leq 4.25.\]

(ii) Path 2: \emph{charge $\rightarrow$ drive $\rightarrow$ empty.} For this path, the \emph{empty} state is reached within the first cycle, 
implying that \(x\) must not exceed \(9\). Starting from \(x = 1\), this limits the duration in \emph{charge} to at most \(2\) time units, 
since \(x = 1 + 4 \cdot t = 9\) for $t=2$. Because the second random clock is never activated along this path, it remains unbounded. Thus,
\[c_0 \in [0, 2], \quad
  c_1 \in [0, \infty].\]

(iii) Path 3: \emph{charge $\rightarrow$ full $\rightarrow$ drive $\rightarrow$ charge $\rightarrow$ drive $\rightarrow$ empty.} In this sequence, the system must first reach the \emph{full} location. Starting at \(x = 2\) and charging at rate \(6\), it takes \(\tfrac{4}{3}\) time units to reach \(x = 10\), since \(x = 2 + 6 \cdot t = 10\) for $t = \tfrac{4}{3}$. Due to the chosen starting value and rate, this is the minimum time required to reach \emph{full}. After the subsequent \emph{drive} phase, \(x\) lies between \(1\) and \(4\). If \(x = 1\), the system can remain in \emph{charge} for at most \(2\) time units before reaching \(x = 9\), ensuring that the \emph{empty} goal remains attainable. This is expressed by the upper bound on $c_1$.

Although the time spent in \emph{full} during the first cycle does not directly affect \(x\), the total duration is bounded by \(100\). Therefore, the time spent in \emph{charge} and \emph{full} (i.e., contributing to \(c_0\)) restricts how long \(c_1\) can run without exceeding this global bound. During the second cycle, in addition to the time required in drive to deplete the current battery state, for each time unit spent in \emph{charge} (with rate \(4\)), \(\tfrac{4}{3}\) time units are needed in \emph{drive} (with rate \(3\)) in order to be able to reach \emph{empty}, yielding a total of \(1 + \tfrac{4}{3} = \tfrac{7}{3}\) time units $t$ per unit of $c_1$. This dependence between both charging durations is expressed as $c_0 + \tfrac{14}{6} c_1 \leq \tfrac{290}{3}$. Thus,
\[c_0 \in \bigl[\tfrac{4}{3}, \tfrac{290}{3}\bigr], \quad
  c_1 \in [0, 2], \quad
  c_0 + \tfrac{7}{3} c_1 \leq \tfrac{290}{3}.\]

 \end{appendix}

\else
\fi

\bibliographystyle{IEEEtran}
\bibliography{references}

\end{document}